\documentclass[letterpaper,twocolumn,10pt]{article}
\usepackage{usenix2019_v3}

\usepackage{booktabs} 
\usepackage[english]{babel}
\usepackage{xcolor}
\usepackage{xspace}
\usepackage{colortbl}
\usepackage{pifont}
\usepackage{subfigure}
\usepackage{subcaption}
\usepackage{graphicx}
\usepackage{booktabs}
\usepackage{threeparttable}
\usepackage{times}
\usepackage[small, compact]{titlesec}
\usepackage[textfont={it}, font={small,bf}, skip=1pt]{caption}

\usepackage{multirow}
\usepackage{fancybox} 
\usepackage{amsmath}
\usepackage{hhline}
\usepackage{amsthm}
\usepackage{comment}
\usepackage{tabularx}
\usepackage{bbm}
\usepackage{amsfonts}
\usepackage{mathtools}
\usepackage{pythonhighlight}
\usepackage[inline,shortlabels]{enumitem}

\usepackage{hyperref}
\definecolor{SoftOrange}{RGB}{230, 145, 56} 
\definecolor{MutedOrange}{RGB}{200, 130, 70}
\definecolor{darkorange}{rgb}{1.0, 0.55, 0.0}
\definecolor{deepcarrotorange}{rgb}{0.99, 0.41, 0.17}
\definecolor{ppurple}{rgb}{0.6, 0.0, 0.5}
\definecolor{plum}{rgb}{0.56, 0.27, 0.52}
\hypersetup{
  colorlinks=true,      
  linkcolor=blue,       
  citecolor=ppurple,    
  filecolor=cyan,       
  urlcolor=red          
}

\newtheorem{lemma}{Lemma}
\usepackage{titling}
\date{}

\usepackage{placeins}

\definecolor{boxcolor}{gray}{0.9}

\newtheorem{theorem}{Theorem}[section]

\usepackage[linesnumbered,ruled,vlined,scleft]{algorithm2e}
\usepackage{float}
\usepackage{algorithmicx}

\definecolor{cardinal}{rgb}{0.77, 0.12, 0.23}

\SetCommentSty{mycommfont}
\newenvironment{denseitemize}{
\begin{itemize}[topsep=2pt, partopsep=0pt, leftmargin=1.5em]
  \setlength{\itemsep}{2pt}
  \setlength{\parskip}{0pt}
  \setlength{\parsep}{0pt}
}{\end{itemize}}

\definecolor{codegreen}{rgb}{0,0.6,0}
\definecolor{codegray}{rgb}{0.5,0.5,0.5}
\definecolor{codepurple}{rgb}{0.58,0,0.82}
\definecolor{backcolour}{rgb}{0.95,0.95,0.92}

\lstdefinestyle{codestyle}{
    commentstyle=\color{SoftOrange}\slshape,
    keywordstyle=\color{black}\bfseries,
    numberstyle=\color{codegray},
    basicstyle=\ttfamily\mdseries\scriptsize,
    emph={AdaEmbed,emb_agent},
    emphstyle={\color{black}\bfseries},
    breakatwhitespace=false, 
    frame=lines,      
    rulecolor=\color{codegray},  
    breaklines=true,                 
    captionpos=b,                    
    keepspaces=true,                 
    numbers=left,                    
    numbersep=5pt,                  
    showspaces=false,                
    showstringspaces=false,
    showtabs=false,                  
    tabsize=2
}
\definecolor{eclipseStrings}{RGB}{42,0.0,255}
\definecolor{eclipseKeywords}{RGB}{127,0,85}
\colorlet{numb}{magenta!60!black}

\lstdefinelanguage{json}{
    basicstyle=\scriptsize\ttfamily,
    commentstyle=\color{eclipseStrings}, 
    stringstyle=\color{eclipseKeywords}, 
    emph={AdaEmbed,emb_agent},
    emphstyle={\color{black}\bfseries},
    numbers=left,
    numberstyle=\scriptsize,
    stepnumber=1,
    numbersep=8pt,
    showstringspaces=false,
    breaklines=true,
    frame=lines,
    string=[s]{"}{"},
    comment=[l]{:\ "},
    morecomment=[l]{:"},
    literate=
        *{0}{{{\color{numb}0}}}{1}
         {1}{{{\color{numb}1}}}{1}
         {2}{{{\color{numb}2}}}{1}
         {3}{{{\color{numb}3}}}{1}
         {4}{{{\color{numb}4}}}{1}
         {5}{{{\color{numb}5}}}{1}
         {6}{{{\color{numb}6}}}{1}
         {7}{{{\color{numb}7}}}{1}
         {8}{{{\color{numb}8}}}{1}
         {9}{{{\color{numb}9}}}{1}
}

\def\ie{{i.e.}}
\def\eg{{e.g.}}

\def\inline1x{Model-S}

\def\name{JITServe\xspace}
\def\algname{\textit{GMAX}\xspace}

\usepackage{tikz}

\newcommand*\blackcircled[1]{\tikz[baseline=(char.base)]{
            \node[shape=circle,fill,inner sep=1pt] (char) {\textcolor{white}{#1}};}}
\algnewcommand{\LeftComment}[1]{\Statex \(\triangleright\) #1}

\begin{document}
\title{\fontsize{15}{15}{\textbf{\name: SLO-aware LLM Serving with Imprecise Request Information}}}

\author{Wei Zhang$^{1}$\thanks{indicates equal contribution.}, Zhiyu Wu$^{1}$\footnotemark[1], Yi Mu$^{1}$, Rui Ning$^{2}$, Banruo Liu$^{1}$, Nikhil Sarda$^{3}$, Myungjin Lee$^{4}$, Fan Lai$^{1}$
\\
\normalsize\textit{$^1$University of Illinois Urbana-Champaign $\quad$ $^2$Unaffiliated $\quad$ $^3$Google $\quad$ $^4$Cisco Research}}


  
\maketitle

\begin{abstract}
The integration of Large Language Models (LLMs) into applications ranging from interactive chatbots to multi-agent systems has introduced a wide spectrum of service-level objectives (SLOs) for responsiveness. These include latency-sensitive requests emphasizing per-token latency in streaming chat, deadline-sensitive requests requiring rapid full responses to trigger external tools, and compound requests with evolving dependencies across multiple LLM calls. Despite---or perhaps, because of---this workload diversity and unpredictable request information (e.g., response lengths and dependencies), existing request schedulers have focused on aggregate performance, unable to ensure application-level SLO needs.

This paper presents \name, the first SLO-aware LLM serving system designed to maximize \emph{service goodput} (e.g., the number of tokens meeting request SLOs) \emph{across diverse workloads}. \name novelly schedules requests using imprecise request information and gradually relaxes this conservatism by refining request information estimates as generation progresses. It applies a \emph{grouped margin goodput maximization} algorithm to allocate just enough serving bandwidth to satisfy each request's SLO \emph{just-in-time} (JIT), maximizing residual capacity for others, while deciding the composition of requests in a batch to maximize efficiency and goodput with provable guarantees. Our evaluation across diverse realistic workloads, including chat, deep research, and agentic pipelines, shows that JITServe improves service goodput by 1.4$\times$-6.3$\times$, alternatively achieving 28.5\%--83.2\% resource savings, compared to state-of-the-art designs.

\end{abstract}
\section{Introduction}
\label{sec:intro}

As large language models (LLMs) enable language-driven interaction between humans and intelligent agents, modern applications increasingly go beyond conventional chatbot scenarios like ChatGPT. They often integrate LLMs with external tools (\eg, AI-assisted coding platforms~\cite{amazonQ} and autonomous web agents~\cite{webpolot_aaai25}), making it crucial to ensure LLM responsiveness (\eg, avoiding external system timeouts~\cite{llm_sys_mobile_arxiv24} or degraded user experience~\cite{alibaba-slo}). Increasingly, applications issue dependent LLM requests to enhance problem-solving, such as response aggregation in test-time scaling~\cite{test-scaling-arxiv25} or coordination in multi-agent systems (MAS)~\cite{gaia-mas}, introducing larger generation token volumes\footnote{A token is a basic unit of text processed or generated by an LLM.} and evolving request dependencies.

The ever-expanding landscape of LLM applications and user bases has introduced increasingly diverse service-level objectives (SLOs) for request responsiveness. 
Our analysis of millions of LLM requests across real-world applications---corroborated by user studies with hundreds of LLM users and extensive discussions with service providers---reveals that requests fall into three dominant patterns (\S\ref{sec:llm-application-intro}):  
(i) \emph{Latency-sensitive requests}: prioritize per-token latency metrics such as time-to-first-token (TTFT) and time-between-tokens (TBT), as in interactive chatbots where incremental streaming directly impacts user experience~\cite{adaserve-arxiv25, alibaba-slo};  
(ii) \emph{Deadline-sensitive requests}: require low end-to-end latency (\ie, E2EL like time-to-last-token) to return a complete response quickly, as seen in cloud AIOps~\cite{llm-cloud}, batch-processing APIs~\cite{Gemini}, or agent interactions that trigger external tools~\cite{llm-monitoring};  
(iii) \emph{Compound requests}: consist of multiple dependent LLM calls to complete a task, for example in MAS~\cite{autellix-arxiv25}, hierarchical reasoning~\cite{tot-neurips-23}, or planning workflows like deep research.  
The latter two categories often interact with external tools or agents, making it critical that their E2EL meets application-level deadlines to avoid downstream timeouts and ensure system reliability~\cite{confucius-sigcomm25}.

As SLOs directly capture application-level performance needs, maximizing service goodput (\ie, effective service throughput like the total number of tokens that meet request SLOs) is essential for both users and service providers. However, deploying dedicated clusters for each SLO workload is impractical due to prohibitive costs and the wide variability of SLO requirements, even among requests of the same type. For example, latency-sensitive chat requests exhibit different TBT requirements depending on user reading speeds~\cite{alibaba-slo, streaming-uist25}, while deadline-sensitive and compound requests in cloud AIOps workflows impose heterogeneous E2EL constraints depending on task urgency and downstream operators (e.g., triggering remediation versus updating monitoring dashboards)~\cite{confucius-sigcomm25, llm-monitoring}.

In the face of growing diversity of workloads and request unpredictability (\eg, response lengths and runtime dependencies), existing LLM serving systems remain misaligned with service goodput. Instead, they typically optimize for aggregate serving throughput~\cite{sarathi-serve-osdi24}, average request completion time~\cite{autellix-arxiv25}, or latency-sensitive requests only~\cite{streaming-uist25}, which we prove can yield arbitrarily poor service goodput (Appendix~\ref{app:prove-prior}). For example, producing a response in 0.5 seconds rather than the expected 5 seconds provides little application benefit if a downstream tool requires 1 minute to execute, yet the excess ``bandwidth'' consumed could have been allocated to requests with more stringent SLOs.

This paper introduces \name, an SLO-aware serving system designed to maximize service goodput across diverse LLM workloads. At its core, \name employs a Just-in-Time (JIT) scheduling principle: it leverages imprecise request information (\eg, response length and dependency) to make initial scheduling decisions, and progressively refines these estimates as generation unfolds. \name then allocates just enough serving bandwidth (\ie, the number of tokens required to generate within a time frame) to satisfy each request's SLO, maximizing the residual capacity for others. \name enables aligning application-level SLO needs with the underlying LLM execution (\eg, vLLM~\cite{vllm-git}) with only a few lines of code modification to existing serving stacks (\S\ref{sec:overview}).

\name addresses two key request uncertainties to guide scheduling. First, while predicting exact response lengths is notoriously error-prone (\S\ref{sec:design-challenges}), we find that estimating an upper bound is both more reliable and more efficient. This upper bound naturally translates into a conservative estimate of the maximum serving bandwidth required. 
To this end, \name leverages a lightweight quantile regression forest (QRF) method~\cite{qclimb-nsdi24} to estimate this bound, and incrementally refines predictions as more response tokens are generated. Second, \name captures evolving execution dependencies in compound requests using \emph{pattern graphs}, where nodes and edges represent LLM/tool inputs, outputs, and their dependencies. By performing incremental graph matching, it identifies prior requests with similar execution patterns, enabling informed group scheduling that mitigates per-stage stragglers (\S\ref{subsec:request-analyzer}).

Beyond uncertainties, request scheduling in LLM serving introduces a unique two-dimensional scheduling challenge. 
First, \emph{single-request scheduling}: even with complete future information (\eg, request length, dependencies, and arrival times), maximizing goodput by scheduling individual requests is already NP-hard. 
Second, \emph{batch composition}: batching requests of heterogeneous lengths slows down per-layer execution due to uneven sequence loads, a bottleneck that persists even under state-of-the-art kernels such as Flash Decoding~\cite{flashattention-22, flash-decoding}. 
To tackle both dimensions, \name proposes a novel \emph{Grouped Margin Goodput Maximization (\algname)} algorithm. The key idea is to ensure that each request receives its minimum serving bandwidth within every time frame by prioritizing those with high margin goodput, allowing surplus bandwidth to be reclaimed in later frames. At the same time, \algname schedules requests with similar input lengths into batches. This joint strategy maximizes goodput and provides provable scaling and performance guarantees (\S\ref{sec:slo-scheduler}--\S\ref{sec:design-dis}).


We implement \name atop vLLM~\cite{vllm-git}, preserving API compatibility to support a wide range of applications (\S\ref{sec:implementation}). We evaluate \name on diverse LLMs and real-world workloads, including chatbots~\cite{intro-chatgpt}, deep research~\cite{DeepResearch}, and agentic workflows~\cite{autogen-arxiv23}. Our results show that \name improves service goodput by 1.4$\times$--6.3$\times$, achieving 28.5\%--83.2\% resource savings for equivalent goodput, while achieving near-oracle performance (\S\ref{sec:eval}).

In summary, our contributions are:
\begin{denseitemize} 
\item We perform real-world studies of LLM services, including user surveys and discussions with providers, introducing a new characterization of request patterns (\S\ref{sec:background});

\item We design a novel scheduler and the \algname algorithm that estimate, refine, and exploit imprecise request information to maximize goodput with provable performance  (\S\ref{sec:overview}-\S\ref{sec:design});

\item We demonstrate \name's significant improvements in application-level performance across real workloads (\S\ref{sec:eval}). 
\end{denseitemize}

\section{Motivation}
\label{sec:background}

We begin with our real-world studies of LLM requests (\S\ref{sec:llm-application-intro}), which reveal new challenges that motivate our work (\S\ref{sec:design-challenges}).

\subsection{Characterizing LLM Serving Requests}
\label{sec:llm-application-intro}

\begin{figure}[t]
  \centering
  \includegraphics[trim=0 80 0 110,clip,scale=0.33]{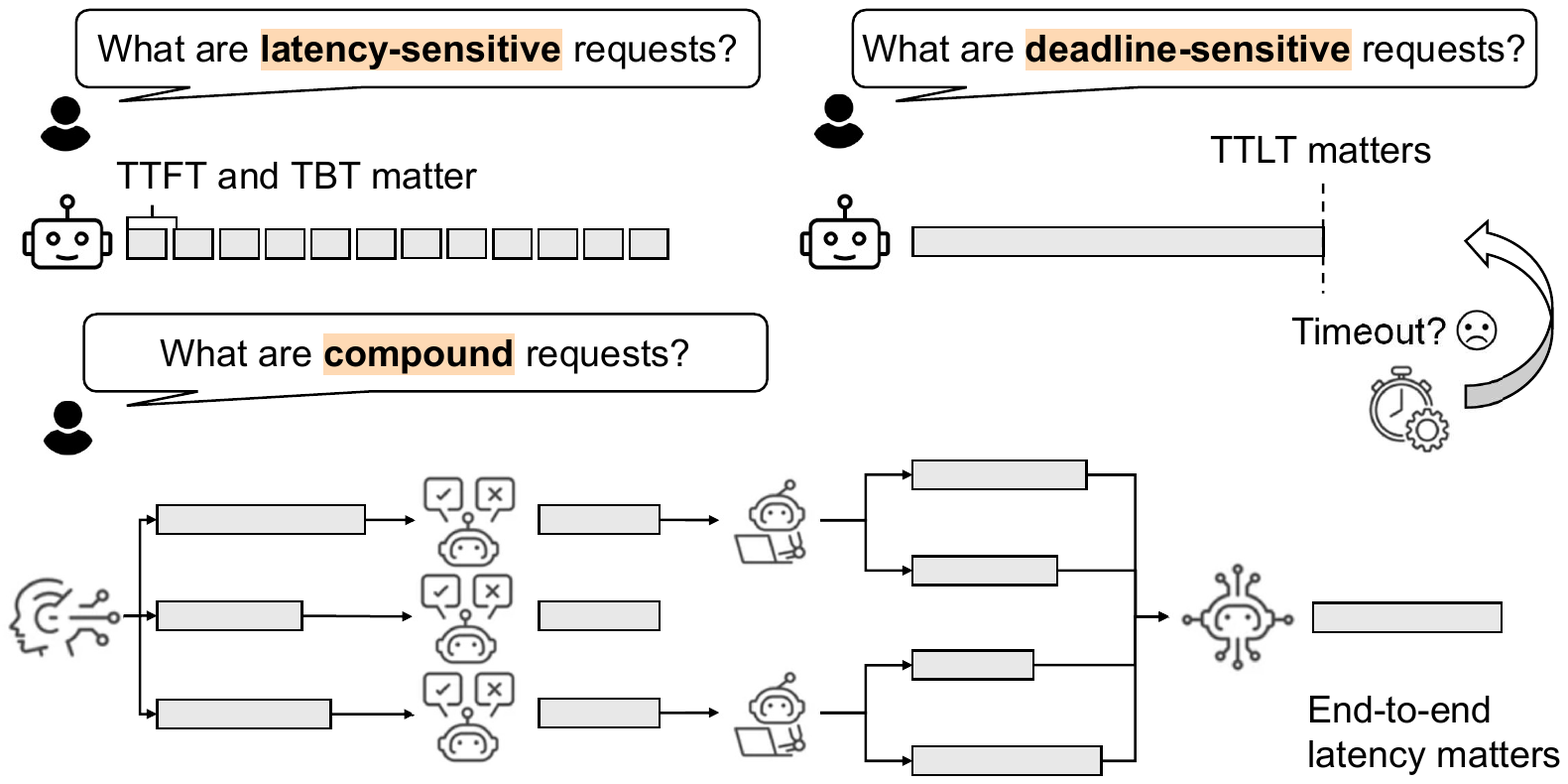}
  \vspace{-.9cm}
  \caption{Illustration of three request patterns.}
  \label{fig:request-patterns}
\end{figure}


As LLMs enable interactive collaboration between human users and AI agents, and proliferate across diverse applications and user bases, optimizing \emph{interaction latency} has become fundamental. To better understand realistic LLM request patterns, we conducted an extensive workload analysis 
covering millions of requests from diverse applications, including chatbots (LMSys Chatbot Arena~\cite{lmsys-chat-23}, WildChat~\cite{wildchat1m}), agentic AI systems (MetaGPT~\cite{metaGPT}, GAIA~\cite{gaia-mas}), and reasoning tasks (deep research~\cite{DeepResearch, ppl-ds}, math reasoning~\cite{tot-neurips-23}). 
To validate and enrich these findings, we engaged in in-depth discussions with two major LLM service providers handling millions of daily requests, and conducted anonymized surveys with over 550 LLM users and developers across six academic and industry organizations. 
Together, our studies reveal that as LLM applications evolve, they introduce new request characteristics and diverse SLOs (Table~\ref{tab:user_survey}).

\begin{table}[t]
\centering
\vspace{0pt}
\begin{tabular}{p{3cm} p{1.2cm} p{1.2cm} p{1.2cm}} 
\toprule
\textbf{LLM Applications} & \textbf{Real-Time} & \textbf{Direct Use} & \textbf{Content-Based}\\
\midrule
Code generation & \textbf{38.1\%} & 30.5\% & 31.4\% \\
Report generation & \textbf{39.1\%} & 36.2\% & 24.7\% \\
Deep research & 38.6\% & \textbf{47.1\%} & 14.3\% \\
Real-time translation & 36.2\% & \textbf{39.9\%} & 23.9\% \\
Batch data processing & 15.6\% & \textbf{49.6\%} & 34.8\% \\
Reasoning task & 28.9\% & \textbf{47.4\%} & 23.7\% \\
\bottomrule
\end{tabular}
\captionof{table}{Our real-world user study reports that users exhibit diverse SLO needs both across and within applications. ``Real-Time'' denotes users who prefer low per-token latency; ``Direct Use'' refers to those demanding for fast full responses; ``Content-Based'' reflects users whose needs vary depending on the specific context.}
\label{tab:user_survey}
\end{table}

\paragraph{Latency-sensitive vs. Deadline-sensitive vs. Compound Requests.}
Our studies show that LLM requests can be increasingly categorized into three key patterns (Figure~\ref{fig:request-patterns}):

\begin{denseitemize}

\item \emph{Type 1: Latency-sensitive Requests.}  They generate responses consumed in a streaming fashion. Representative applications include ChatGPT-style web services, AI-powered customer support~\cite{ai-customer-service}, and real-time speech-to-text services~\cite{openai-tts}. For such requests, it is critical to maintain a content delivery rate (\ie, TTFT and TBT) that matches or exceeds the user's consumption pace (e.g., reading speed) to ensure a smooth interactive experience~\cite{streaming-uist25, alibaba-slo}. 

\item \emph{Type 2: Deadline-sensitive Requests.}  
They require the full response (\ie, E2EL) to be generated within a specified deadline (\eg, to prevent downstream tool timeouts). This pattern arises in applications such as agent interactions that invoke external tools (e.g., for cloud AIOps~\cite{llm-cloud, llm-monitoring}), data cleaning~\cite{llmclean}, large codebase generation~\cite{code-gen-arxiv24}, and batch processing APIs at OpenAI and Gemini~\cite{Gemini}. 

\item \emph{Type 3: Compound Requests.}
These involve multiple dependent LLM calls, often forming graph-structured execution dependencies~\cite{parrot-osdi24, ayo-asplos25, autellix-arxiv25}, with the requirement that the entire end-to-end generation (\ie, E2EL of finishing all requests) completes within the deadline. Examples include reasoning tasks using test-time scaling~\cite{tot-neurips-23}, multi-agent workflows~\cite{autellix-arxiv25}, hierarchical planning scenarios such as deep research~\cite{plan-and-solve, deep-searcher}, and reinforcement learning from human feedback pipelines where multiple responses must be generated per prompt~\cite{rlhf-neurips22}. 

\end{denseitemize}

\paragraph{Need for Accommodating Diverse SLO Requirements.}
Practical LLM workloads often involve mixed request types, and even a single request may transition across types during execution. 
For example, our user studies find that in multi-step reasoning tasks, the initial ``thinking'' phase is often deadline-sensitive---users expect internal reasoning to complete within seconds to avoid perceived stalls. 
Once the generation transitions to producing the final response, the workload becomes latency-sensitive, where smooth TBT is critical for interactive reading. 
This observation has been corroborated by other user-experience studies~\cite{user-experience-llm-arxiv24, streaming-uist25} and our discussions with service providers. 
Further, certain requests may not impose explicit SLOs (e.g., synthetic data generation~\cite{syn-data-gen-arxiv23}), yet do not want to suffer starvation. 

Beyond workload and application heterogeneity, SLO requirements also vary across users. 
As shown in Table~\ref{tab:user_survey}, 30.5\% of users prefer minimizing E2EL in code generation for rapid testing and direct execution, whereas 38.1\% favor streaming code delivery to facilitate interactive reading and comprehension. 
Even within streaming use cases, users exhibit different reading speeds, translating into heterogeneous TBT requirements~\cite{andes-arxiv24, alibaba-slo, streaming-uist25}.  
Batch APIs provide differentiated response guarantees across pricing tiers, leading to distinct deadline constraints~\cite{openai-batch-API,OpenaiAPIPricing}.  
Agentic workflows also diverge in their downstream integrations (\eg, triggering automated remediation or updating monitoring dashboards), resulting in varied E2EL requirements~\cite{benchmarking-agentic-arxiv24}.  

A naive approach to request diversity is to dedicate clusters to each workload type. However, this is both cost-prohibitive and insufficient, as SLO needs vary even among requests of the same type. Worse, request types may evolve during execution (\eg, from ``reasoning'' to ``streaming'' phases), making migration (\eg, KV cache) across clusters costly.



\subsection{Challenges and Limitations of Existing Solutions}
\label{sec:design-challenges}

As widely recognized in application-aware networking~\cite{varys-sigcomm14, Karuna}, computing~\cite{erlang-eurosys24}, and storage~\cite{app-storage-fast19, extmem-atc24}, SLOs directly capture application-level performance needs. Completing requests much faster than their specified demands yields little additional \emph{service goodput}. An effective scheduler should therefore allocate just enough serving ``bandwidth'' to satisfy each request's SLO, meeting application requirements while maximizing residual capacity for other requests or enabling substantial resource savings. Unfortunately, modern LLM serving systems face unique challenges and fundamental inefficiencies under this growing workload diversity.

\begin{figure}[t]
  \centering
  {
    \subfigure[Varying number of LLM calls. \label{fig:config-xs}]{\includegraphics[width=0.49\linewidth]{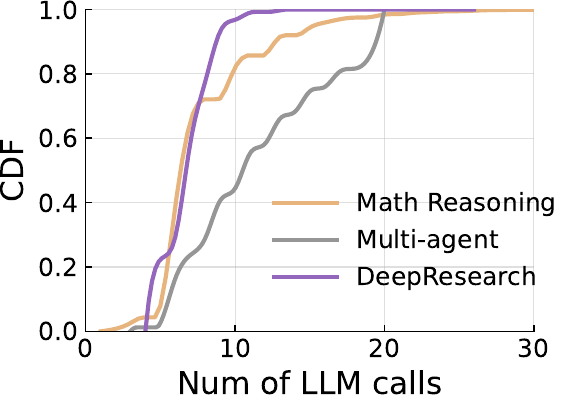}}
    \subfigure[Large prediction deviations. \label{fig:config-xl}]{\includegraphics[width=0.49\linewidth]{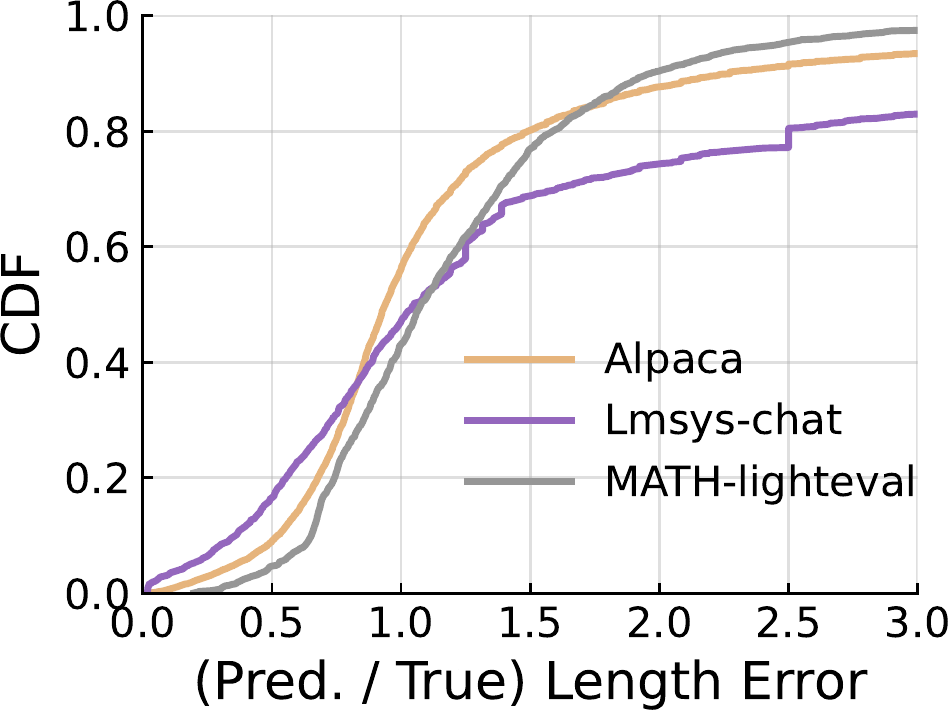}}
  }
  \caption{(a) LLM workloads increasingly involve varying numbers of subrequests (LLM calls) in a request. (b) Predicting response length remains highly inaccurate.}
  \label{fig:autoregressive-uncertainty}
\end{figure}

\paragraph{Pervasive Request Uncertainties.} 
LLM request scheduling must contend with multiple sources of uncertainty. 
First, request arrival patterns in online serving can fluctuate sharply, with load variations of up to $5\times$ within minutes~\cite{dynamollm-isca24}.  
Second, modern LLM workloads often exhibit complex and unpredictable execution dependencies. 
As shown in Figure~\ref{fig:config-xs}, applications such as multi-agent workflows~\cite{autogen-arxiv23}, test-time scaling for reasoning~\cite{tot-neurips-23}, and deep research tasks introduce highly variable numbers of LLM invocations, often unknown a priori due to reflective reasoning or adaptive exploration (\eg, until reaching sufficient confidence~\cite{dr-github}). 

Finally, even when the dependency structure is known, predicting response lengths---especially for downstream requests whose inputs are not yet available but are required to ensure E2EL---is highly impractical, further challenged by probabilistic token sampling and self-reflection dynamics in response generation~\cite{self-refine-neurips23}. 
As shown in Figure~\ref{fig:config-xl}, even with the full prompt input provided, both self-prediction (e.g., Gemini estimating its own output length) and fine-tuned predictors (e.g., BERT- or Llama3-based models~\cite{userve-atc24}) exhibit substantial length prediction errors.

\begin{figure}[t]
  \centering
  \includegraphics[width=\linewidth]{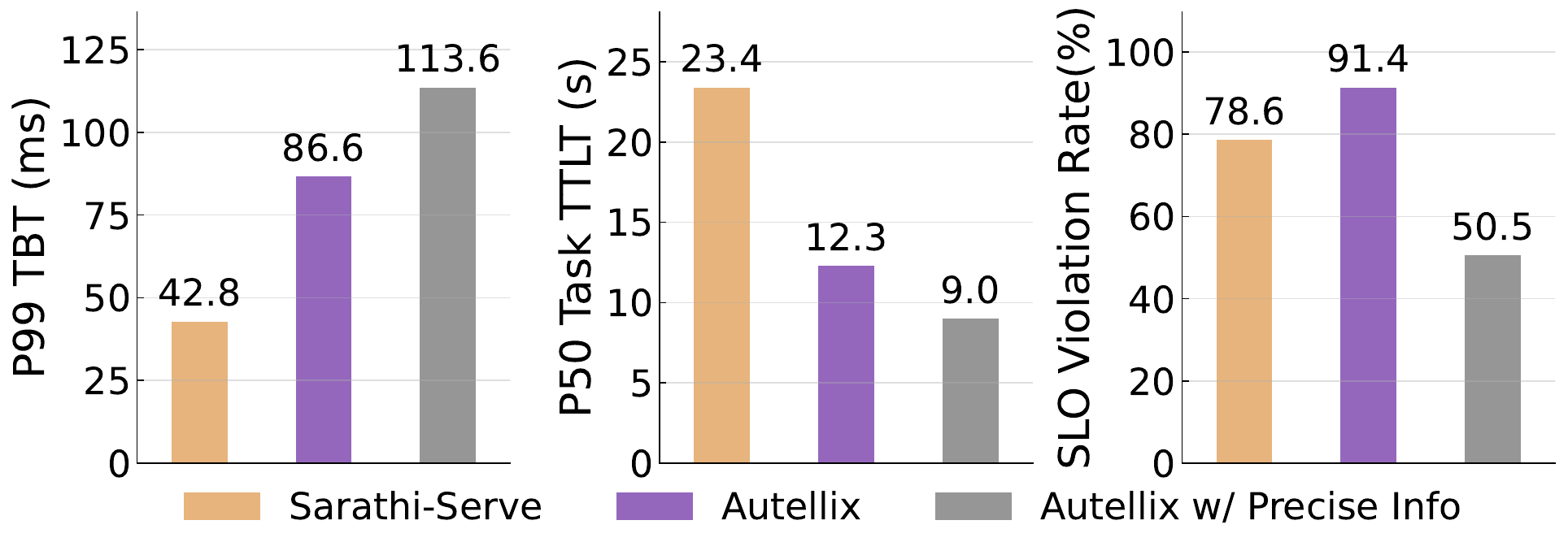}
  \caption{Existing advances face significant performance drops due to growing LLM request diversity.}
  \label{fig:limitation_existing_solutions}
\end{figure}

  



\paragraph{Misaligned Service Goodput and Inefficiency in Existing Solutions.}
State-of-the-art schedulers fail to generalize under increasing workload and SLO diversity.  
First, they primarily optimize aggregate metrics, such as minimizing mean E2EL via Shortest-Job-First (SJF) variants with predicted length ranking~\cite{length-predict-neurips24} or Least-Attained Service (LAS) First as in Autellix~\cite{autellix-arxiv25}.  
However, we theoretically prove that they can lead to arbitrarily poor goodput (Appendix~\S\ref{app:prove-prior}): even if the mean improves, many requests still miss their SLOs, leading to service unpredictability, cascading violations across dependent tasks, and overclaiming resources to sustain service.  
As concrete evidence, Figure~\ref{fig:limitation_existing_solutions} shows that while Autellix improves average E2EL compared to Sarathi-Serve~\cite{sarathi-serve-osdi24}, it suffers from higher and over 90\% SLO violation rates.

Second, perhaps due to the challenges of request uncertainty, existing serving optimizations that consider user experience are restricted to latency-sensitive workloads (e.g., Sarathi-Serve), since TTFT and TBT primarily depend on known input lengths.  
Yet, as shown in Figure~\ref{fig:limitation_existing_solutions}, Sarathi-Serve achieves low TBTs but performs poorly for deadline-sensitive requests (\eg, large TTLTs), resulting in high SLO violations.  
Finally, even within their intended design regimes, these schedulers fall substantially short of the oracle baseline with perfect knowledge of request lengths and dependencies: Figure~\ref{fig:limitation_existing_solutions} demonstrates that Autellix achieves 41\% higher SLO attainment rates when provided with precise information.

Addressing these limitations is critical for both LLM service users and providers, calling for a new scheduler that explicitly aligns LLM execution with application-level needs and satisfies three essential properties:  
\begin{denseitemize}  
\item \emph{Generalizability}: Support diverse request types (\eg, latency-, deadline-, and compound requests) and SLO requirements, ensuring predictable SLO satisfaction without collecting intrusive request and application information. 

\item \emph{Goodput Efficiency}: Maximize service goodput and resource utilization for providers, while remaining robust to runtime dynamics with provable performance guarantees.  

\item \emph{Deployability and Scalability}: Integrate seamlessly with existing serving stacks and provide strong scaling capabilities (e.g., extending to multiple concurrent models).  
\end{denseitemize}

\section{\name Overview}
\label{sec:overview}

This paper introduces \name, an SLO-aware LLM request scheduler that generalizes across diverse workloads and SLOs, maximizing service goodput (\eg, the number of useful tokens delivered) and resource utilization for both LLM users and providers under imprecise request information. \name provides provable guarantees and achieves empirically near-oracle performance, complementing existing serving infrastructure with only a few lines of code change in APIs. 

\paragraph{Design Space.}
\name targets practical serving deployments where requests arrive online and completed ones exit, covering diverse LLM workloads and SLOs. We adopt the widely used notion of goodput~\cite{bentoml-goodput}:  
\begin{denseitemize}
\item \emph{Latency-sensitive requests}: measured as the number of tokens delivered within the expected \emph{timeline}~\cite{streaming-uist25, alibaba-slo}. Specifically, token $i$ counts toward goodput if it finishes by $TTFT_{SLO} + i \times TBT_{SLO}$. 
\item \emph{Deadline-sensitive requests}: measured as the total number of tokens, including input and output, if the request completes before its deadline; zero otherwise~\cite{alibaba-slo}.  
\item \emph{Compound requests}: measured as the total number of tokens across all subrequests if the final generation completes by the E2EL deadline; zero otherwise~\cite{autellix-arxiv25}.  
\end{denseitemize} 

\name is agnostic to the specific definition of goodput and operates directly over the metric provided by the service provider. For example, the provider can define the final subrequest in compound requests as latency-sensitive requests in the goodput objective function. It also accommodates non-SLO requests (\eg, best-effort requests) by assigning a default completion deadline to avoid starvation (\S\ref{sec:slo-scheduler}). We further validate that \name consistently improves various goodput objectives (\S\ref{eval:e2e}), such as request-level goodput like maximizing the number of requests that meet their SLOs~\cite{distserve-osdi24}.

At its core, \name employs a Just-in-Time (JIT) scheduling strategy to be conservative yet adaptive under runtime uncertainties. Instead of assuming precise knowledge or ignoring uncertain information, \name initially estimates quantile-based upper bounds for response lengths and request dependency graphs, allocating conservative bandwidth to prevent SLO violations. As response generation progresses, \name refines these estimates, gradually relaxing bandwidth allocations to maximize residual capacity for other requests.

\begin{figure}[t]
  \centering
  \includegraphics[width=\linewidth]{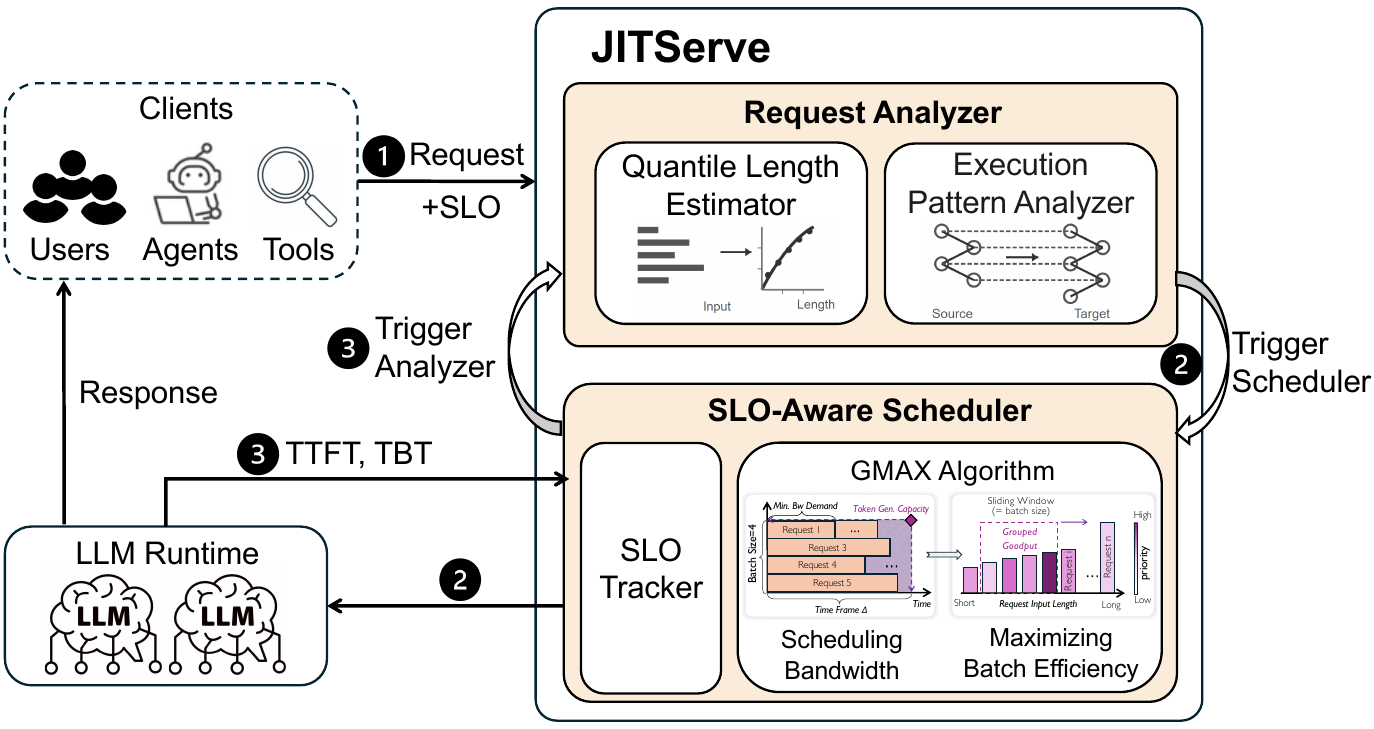}
  \caption{\name system overview and workflow.}
  \vspace{.2cm}
  \label{fig:sys-arc}
\end{figure}

\paragraph{System Workflow.}
As illustrated in Figure~\ref{fig:sys-arc}, \name operates as a middleware layer that aligns application-level performance needs with underlying execution backends (e.g., vLLM~\cite{vllm-git}).  
\blackcircled{1} Upon request arrival, along with its SLO (specified by users~\cite{andes-arxiv24} or developers~\cite{adaserve-arxiv25}), the Request Analyzer estimates key request information, including an upper bound on output length and execution dependencies (e.g., via graph matching to historical request patterns).  
\blackcircled{2} The SLO-Aware Scheduler determines the minimum number of tokens each request should generate within a time frame, prioritizing those with higher margin goodput while grouping requests with similar input lengths into batches to maximize per-iteration batch execution speed.  
\blackcircled{3} Leveraging the SLO Tracker, which monitors actual generation speeds and continuously updates estimates from the Request Analyzer, the scheduler efficiently makes admission and preemption decisions.

\section{\name Design}  
\label{sec:design}

SLO-aware LLM serving must address three fundamental challenges: 
(1) \emph{Generalizability}: proactively estimating and refining request information during execution to meet diverse SLOs (\S\ref{subsec:request-analyzer}), which informs (2) \emph{Goodput Efficiency}: scheduling individual requests and their batch compositions to maximize service goodput (\S\ref{sec:slo-scheduler}), and
(3) \emph{Deployability}: adapting to diverse deployment considerations, such as fairness and strong scaling capabilities (\S\ref{sec:design-dis}).
We next describe how \name addresses them in real time.

\subsection{Request Analyzer: Refining Imprecise Information}
\label{subsec:request-analyzer}

Predicting request information is notoriously difficult, especially for downstream requests without knowing their prompt inputs, yet we prove that scheduling without it---e.g., using traditional Least-Attained Service~\cite{autellix-arxiv25} or Earliest-Deadline-First policies---can lead to arbitrarily poor service goodput due to the misalignment between optimizing for aggregate performance and meeting individual SLOs (Appendix~\ref{app:prove-prior}). 
Our key insight is that LLM responses are generated autoregressively over hundreds of decoding iterations, enabling a middle-ground approach between assuming full information and assuming none. Specifically, we leverage imprecise but actionable, continuously refinable estimates to inform scheduling.  
To this end, \name employs two key techniques using the information already available in serving: 
(1) \emph{quantile-based upper-bound prediction of response length}, and 
(2) \emph{dynamic pattern-graph matching} based on history.

\paragraph{Estimating Response Length Upper-bound.}
Determining just enough serving bandwidth to meet each request's SLO requirements requires knowing its response length. Underestimation risks SLO violations as deferring long-response requests by mistake will miss deadlines, while overestimation wastes capacity due to overclaiming bandwidth. 

\begin{figure}[t]
  \centering
  {
    \subfigure[Estimation overhead. \label{fig:rf-latency}]{\includegraphics[width=0.49\linewidth]{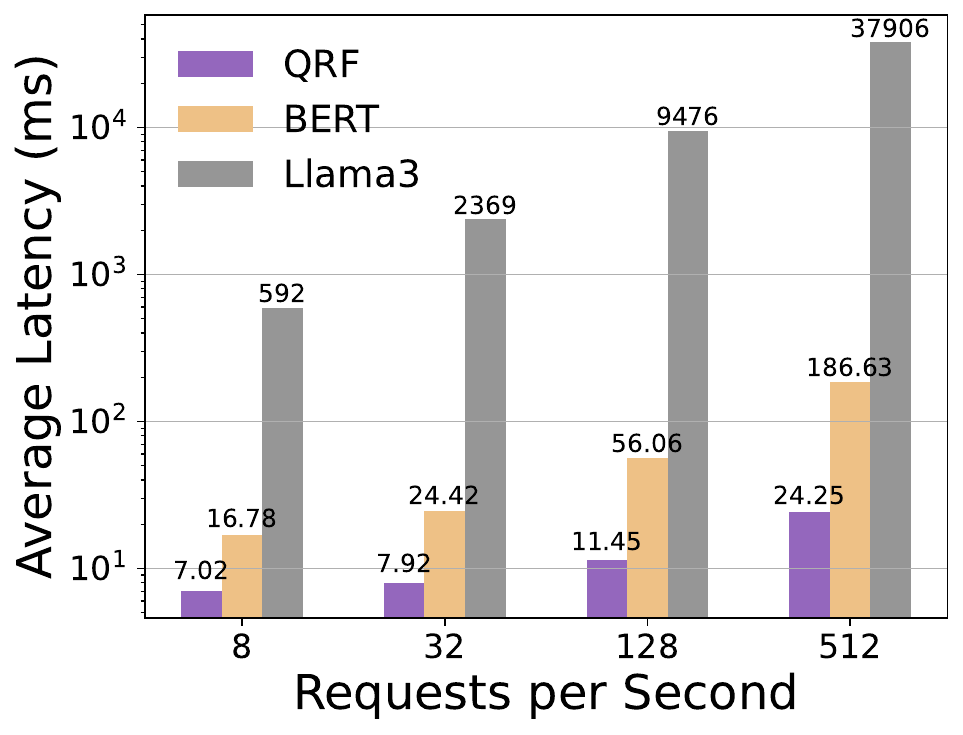}}
    \subfigure[Estimation accuracy. \label{fig:rf-online}]{\includegraphics[width=0.49\linewidth]{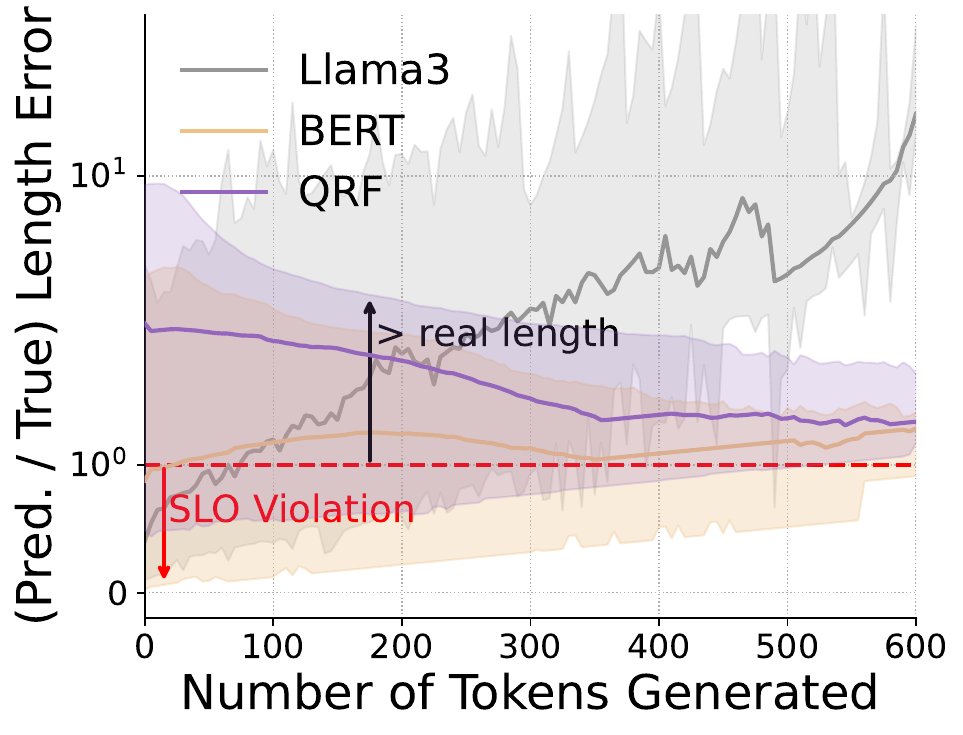}}
  }
  \caption{(a) Average prediction latency. (b) QRF achieves better upper-bound prediction than tuned BERT and LLama3-based predictor over time. The shaded area shows P5-P95 distribution, and the red line represents the ground truth (ratio = 1).}
  \label{fig:qrf-overall}
\end{figure}

To balance these risks, \name conservatively overestimates response length yet progressively refines the estimate over generations. 
It uses a Quantile Regression Forest (QRF) model~\cite{QRF} for response length prediction. Compared to range classification approaches~\cite{userve-atc24} that require predetermined buckets, QRF customizes quantile intervals in predictions for each specific request, making it more adaptable across diverse prompts and models. We leverage QRF to output a high-quantile bound on response length given a request's prompt, which corresponds to the maximum bandwidth requirement. 
The QRF prediction is lightweight:  
as shown in Figure~\ref{fig:rf-latency}, each prediction takes only 7 ms,  7$\times$ faster than a fine-tuned BERT predictor~\cite{userve-atc24}. 
 
Our lightweight design makes online refinement practical: the Request Analyzer augments the prompt with newly generated tokens and periodically re-invokes QRF (\eg, every 50 tokens) to refine the upper bound, thereby progressively reducing conservatism. This provides two additional benefits: (1) it adapts to generation-time variability, where even identical prompts may yield different outputs from the same model (\eg, due to probabilistic token sampling); and (2) it generalizes across models by incrementally incorporating model-specific responses into predictions, thereby improving accuracy. 
Even disregarding BERT's substantial overhead, Figure~\ref{fig:rf-online} shows that both fine-tuned BERT and Llama3 prediction models often underestimate response lengths, risking frequent SLO violations. In contrast, QRF produces reliable upper-bound estimates that relax as generation progresses.

\paragraph{Estimating Dependency with Pattern-Graph Matching.}
Beyond uncertainty at the individual request level, many workloads exhibit graph-structured execution (\eg, deep research~\cite{deep-searcher}), where nodes represent LLM or tool invocations and edges encode their dependencies. Moreover, these dependency structures may evolve dynamically (\eg, to achieve sufficient response confidence~\cite{deep-searcher}), making it difficult to satisfy SLOs (\eg, E2EL), since the response generation of deeper-stage requests depends on unknown parents' outputs (\S\ref{sec:design-challenges}).



\begin{figure}[t]
  \centering
  \includegraphics[width=\linewidth]{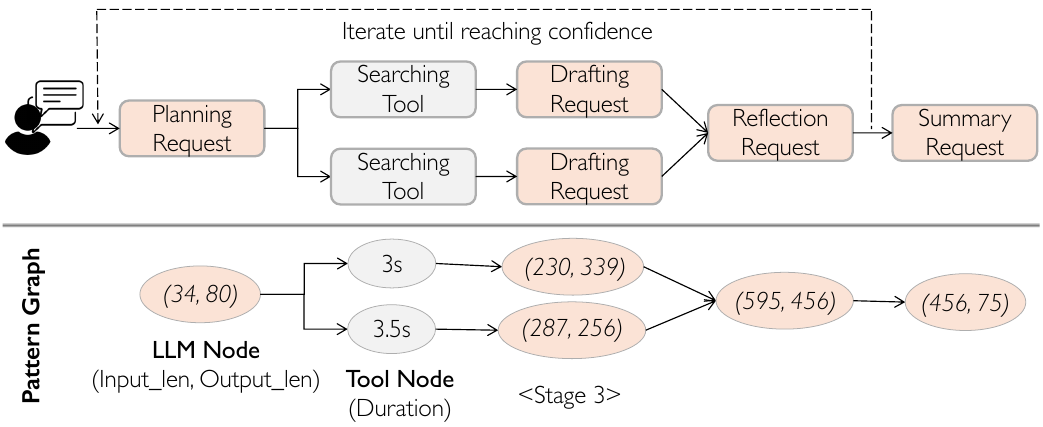}
  \caption{Example of Pattern Graph consisting of five stages. LLM node weighted by (input\_len, output\_len), tool node weighted by execution time.}
  \label{fig:super-node-illustration}
\end{figure}

Our key insight is twofold: (1) exploit historical requests with structurally similar execution graphs to infer likely dependency patterns, and (2) amortize SLO requirements (\eg, deadlines) across intermediate stages to ensure steady progress, thereby avoiding overly deep planning that introduces significant noise.  
We represent each served request as a primitive \emph{pattern graph}, without needing raw input/output plaintext. As illustrated in Figure~\ref{fig:super-node-illustration}, each node correspond to one LLM or tool invocation, annotated with input/output length (for LLMs), execution time (for tools), and the model/tool identity, while edges capture node dependencies.

As a new request unfolds in response lengths and new invocations, the Request Analyzer incrementally extends its partial graph with newly revealed dependencies, prunes past patterns (graphs) whose \emph{prefix structures} diverge (\eg, invoking a different model/tool at the current stage), and performs similarity matching against the remaining candidates. 
Node and edge similarities are computed using Gaussian-kernel functions~\cite{graph-kernel-20} over their attributes (output lengths for nodes, input lengths for edges), enabling progressively refined pattern matching as more information becomes available.


Once the most similar historical pattern graph is identified, we use it to estimate the cumulative contribution of prior stages relative to overall execution. This enables proportional sub-deadline allocation across stages rather than treating them uniformly.  
Specifically, we compute the accumulated share as
$
\phi(s) = \frac{t_{\leq s}}{t_{\text{total}}},
$
where $t_{\leq s}$ is the accumulated execution time up to stage $s$, and $t_{\text{total}}$ is the total execution time \emph{in the pattern graph}. Intuitively, $\phi(s)$ captures the historical progress made up to stage $s$ as a fraction of the full execution timeline.  
The amortized deadline for stage $s$ in a new request with total deadline $D$ is then set as
$
D_s = \phi(s) \cdot D,
$
ensuring that each stage receives a sub-deadline proportional to its cumulative contribution. We also evaluated alternative formulations (\eg, $t_s/t_{\text{total}}$) and found that our accumulated-share design consistently outperforms them in both analytical modeling and empirical evaluation (Appendix~\ref{app:pattern-graph}), offering greater robustness and accuracy by grouping previous stages' information.

\begin{figure}[t]
  \centering
  
  \includegraphics[trim=0 279 0 0,clip, width=0.98\linewidth]{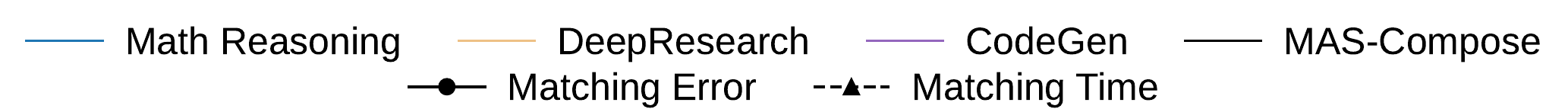}

 \subfigure[Impact of historical repo size.\label{fig:graph-hist}]{\includegraphics[trim=15 0 450 68,clip, width=0.53\linewidth]{Figures/design/combined_error_and_time.pdf}}
\subfigure[Impact of online matching.\label{fig:graph-online-accuracy}]{\includegraphics[trim=540 0 15 68,clip, width=0.45\linewidth]{Figures/design/combined_error_and_time.pdf}}

  \caption{
  \subref{fig:graph-hist} larger historical graph sets reduce matching error while exhibiting sublinear time growth. \subref{fig:graph-online-accuracy} next-stage estimation error decreases as more stage information becomes available. Note that the next-stage estimation error becomes zero when the maximum number of stages is already reached (i.e., $t_s = 0$).}
  \label{fig:graph-match-accuracy}
\end{figure}

To ensure efficiency, we cluster historical pattern graphs offline using a K-medoids mechanism~\cite{k-medoids}, and evict patterns with low reuse frequency (decayed by 0.9 every hour). Each stored pattern graph is compact, typically under 0.2KB. 
As shown in Figure~\ref{fig:graph-hist}, our matching procedure achieves both high efficiency and accuracy: the matching latency remains below 5\,ms even with 500 historical graphs, while accuracy already saturates with such a modest history size. 
This lightweight design enables online matching in real-time serving. 
As shown in Figure~\ref{fig:graph-online-accuracy}, the relative error in next-stage ratio estimation decreases as additional stage information becomes available, demonstrating progressive refinement.



\subsection{SLO-aware Scheduler with \algname Algorithm}
\label{sec:slo-scheduler}

\begin{figure}[t]
  \centering
  \begin{minipage}{0.48\linewidth}
\centering
    \includegraphics[width=\linewidth] 
    {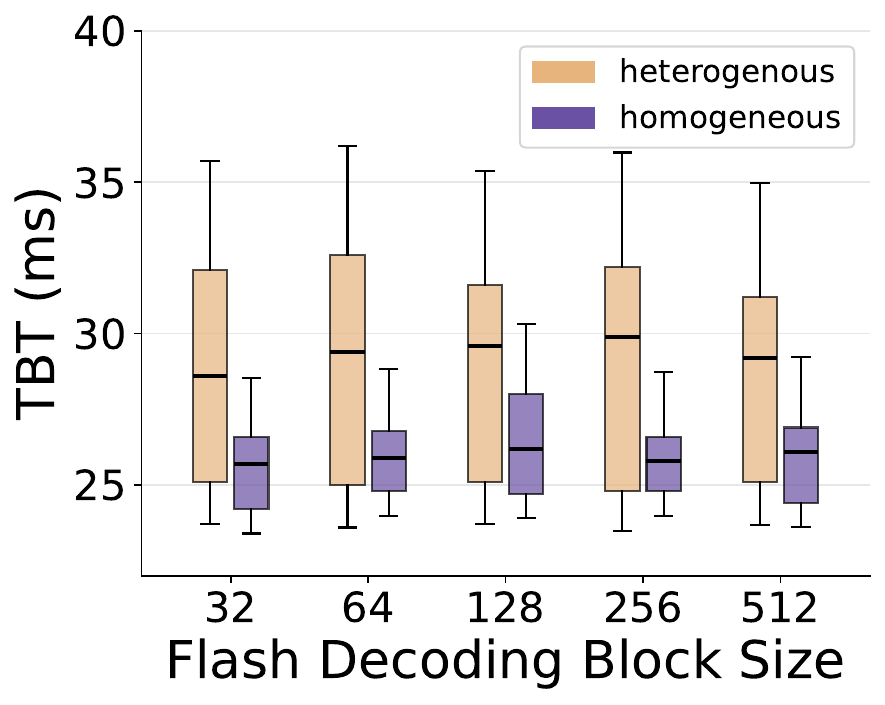}
   \captionof{figure}{Batching requests with heterogeneous lengths slows down execution.}
   \vspace{-0.4em}
   \label{fig:heter-homo}
   \end{minipage}
   \hfill 
   \begin{minipage}{0.48\linewidth}
   \centering
   \includegraphics[width=\linewidth] {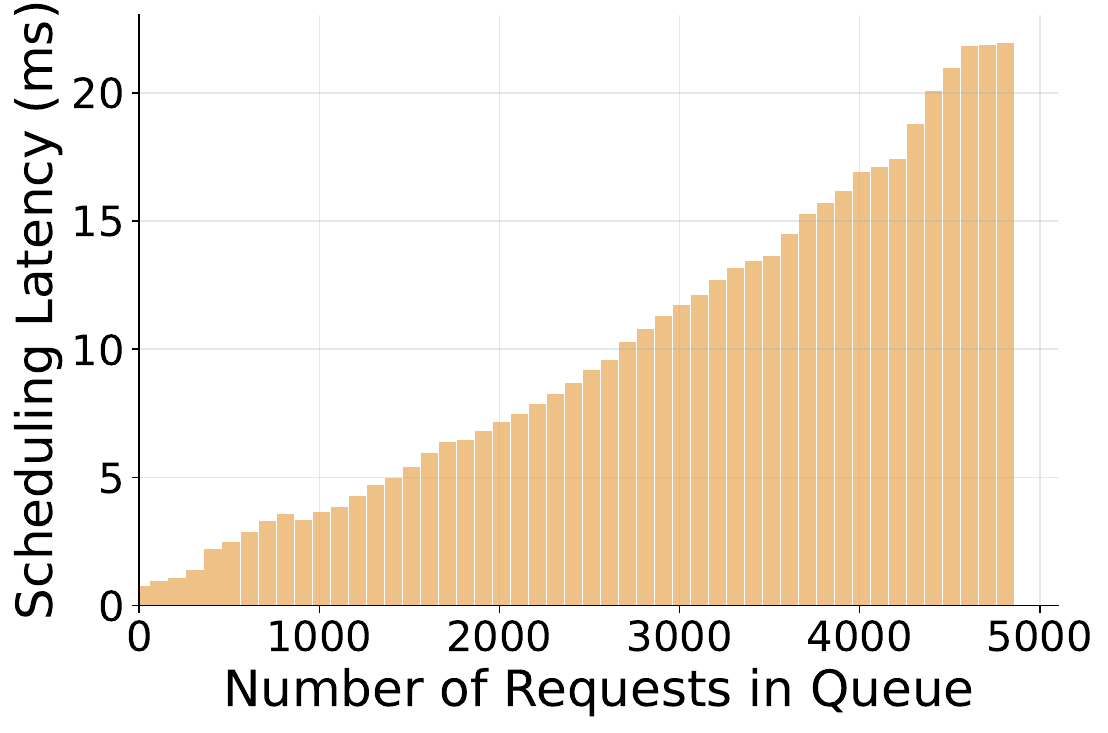}
   \captionof{figure}{\algname scales efficiently to schedule thousands of concurrent requests.}
   \vspace{0.3em}
  \label{fig:scheduling-scaling}
  \end{minipage}
\end{figure}

With imprecise yet continuously refined request estimates, the SLO-aware scheduler aims to maximize service goodput by allocating just enough serving bandwidth (\eg, the minimum generation tokens required within a time frame) to satisfy each request's SLO, thereby preserving residual capacity for others. This creates a unique two-dimensional scheduling challenge that extends beyond traditional scheduling problems.
First, \emph{single-request scheduling}: even under complete future information (\eg, exact response length, dependency, and arrival time), we prove that maximizing goodput by scheduling individual requests is already NP-hard (Appendix~\ref{sec:OPT}).
Second, \emph{batch composition}: LLMs execute requests in batches, but batching requests with heterogeneous input lengths reduces per-token generation speed due to uneven input loads across samples in each model layer's batch execution, distinct from prior continuous batching problems~\cite{orca-osdi22}. As shown in Figure~\ref{fig:heter-homo}, this inefficiency persists even with advanced kernels such as Flash Decoding~\cite{flash-decoding}.

Algorithm~\ref{algo:concord_runtime_class} summarizes how \name addresses the online two-dimensional scheduling challenge with the \emph{Grouped Margin Goodput Maximization} (\algname) algorithm. 
The scheduler first queries the Request Analyzer to determine each request's \emph{minimum serving bandwidth} requirement (Lines~\ref{algo:analyze-start}-\ref{algo:analyze-end}). 
It then prioritizes requests with the highest goodput payoff relative to their bandwidth consumption, while grouping (scheduling) those with similar input lengths into a batch to maximize grouped goodput and 
batching efficiency (Lines~\ref{algo:sliding-start}-\ref{algo:sliding-end}). 
As generation progresses, both the Request Analyzer and the scheduler continuously refine request estimates and update scheduling decisions accordingly (Lines~\ref{algo:online-prediction-start}-\ref{algo:online-prediction-end}).

\begin{algorithm}[t]
  \caption{\algname Algorithm}
  \label{algo:concord_runtime_class}
  \SetKwProg{Class}{Class}{:}{}
  \SetKwProg{Function}{Function}{:}{}

  \Class{\textbf{RequestAnalyzer}}{
    \Function{\textsc{AnalyzeRequest}(req)}{ 
        \label{algo:analyze-start}
      \tcp{Capture minimum serving bandwidth each request needs to meet SLOs}
      $\textit{req}.len\_{\mathrm{rem}} \gets \textsc{{PredictLength}}(\textit{req})$ \\
      $\textit{req}.bw \gets \dfrac{\textit{req}.len\_{\mathrm{rem}}}{\textsc{EstimateRemainingTime}(\textit{req})}$ \\
      $\textit{req}.goodput \gets \textsc{EstimateGoodput}(\textit{req})$ \\
      $\textit{req}.priority \gets \dfrac{\textit{req}.goodput}{\textit{req}.bw}$ \\
      \label{algo:analyze-end}
    }
  }
  \BlankLine
      \Class{\textbf{Scheduler}}{
      \Function{\textsc{Schedule}(batch\_size, cutoff)}{
        $\textit{Q} \gets \textsc{GetRequestQueue()}$ \\
        \ForEach{$req$ in $Q$}{
        \label{algo:online-prediction-start}
            \textsc{AnalyzeRequest}($req$)\\
        }
        \label{algo:online-prediction-end}
        $bp \gets \textsc{BatchPriority}(\textit{Q}, \textit{batch\_size})$
        \BlankLine
        \tcp{Step 1: candidate filtering by priority cutoff $p$}
        $Candidates \gets Q.\textsc{Filter}(req: req.priority \ge bp \times cutoff)$

        \BlankLine
        \tcp{Step 2: group by input length (sliding window)}
        \label{algo:sliding-start}
        $Candidates.\textsc{sort}(req:req.input\_length)$ \\
        $BestGroup \gets \emptyset$,  $max\_score \gets -\infty$ \\
        \ForEach{window $\mathcal{G}$ of size $B$ in Candidates}{
          $score \gets \sum_{r \in \mathcal{G}} r.priority$ \\
          \If{score > max\_score}{
            $BestGroup \gets \mathcal{G}$, $max\_score \gets score$
          }
        }
        \label{algo:sliding-end}
        \Return BestGroup
      }
    }
\end{algorithm}

\begin{figure}[t]
    \centering
    \includegraphics[width=\linewidth]{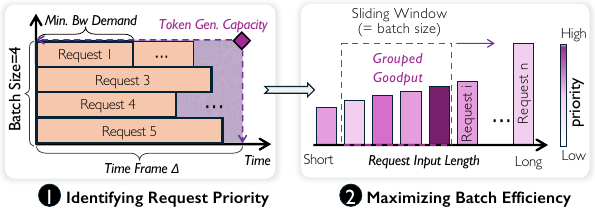}
    \caption{\algname (1) identifies the scheduling priority of each request from its bandwidth demand, and then (2) selects requests to maximize the grouped margin goodput and batch efficiency. }
    \label{fig:bin-packing-sliding-window}
\end{figure}

\paragraph{Capturing Minimum Serving Bandwidth per Request.} 
For each request $r$, the \emph{minimum serving bandwidth} depends on its remaining work (i.e., the remaining response length to generate) and the remaining time budget. Formally, we define:
$
bw(r) = \frac{t_{\text{gen}}(r)}{t_{\text{rem}}(r)},
$
where $t_{\text{gen}}(r) = len_{\text{rem}}(r) \cdot v_{\text{token}}(r)$ is the upper-bound estimate of the remaining generation time, computed as the remaining response length $len_{\text{rem}}(r)$ upper bound times the average per-token generation speed $v_{\text{token}}(r)$. This estimate is conservatively initialized and incrementally refined during generation (\S\ref{subsec:request-analyzer}). $t_{\text{rem}}(r)$ is the remaining time to the request deadline for deadline-sensitive or compound requests. For latency-sensitive requests, explicit SLOs (e.g., $TBT$) already define the per-token service bandwidth. 

Because maintaining a fixed bandwidth throughout a request's lifetime is impractical due to runtime dynamics (e.g., new request arrivals or early completions), \algname amortizes bandwidth allocation over discrete scheduling frames of length $\Delta$. As illustrated in Figure~\ref{fig:bin-packing-sliding-window}, execution is decomposed into consecutive frames. 
Each frame provides a token-generation capacity (the purple shaded area) across the time ($\Delta$) and batch-size dimensions. A request $r$ is represented as a rectangle occupying one batch slot with a frame-level bandwidth of
$
bw_\Delta(r) = \frac{t_{\text{gen}}(r)}{t_{\text{rem}}(r)} \cdot \Delta.
$
For compound requests, both $len_{\text{rem}(r)}$ and $bw_\Delta(r)$ are aggregated across all subrequests within the current stage, since completing a single subrequest does not advance the stage.  

Analogously, we amortize each request's potential goodput:  
$
goodput_\Delta(r) = \frac{goodput(r)}{t_{\text{rem}}(r)} \cdot \Delta
$, 
where $goodput(r)$ denotes the achievable goodput contribution of completing $r$, depending on the developer's SLO specification (\S\ref{sec:overview}). Scheduling thus reduces to efficiently placing request rectangles into the per-frame capacity while maximizing aggregate $goodput_\Delta$. A natural solution is dynamic programming (e.g., $DP(\mathcal{R}, t, B)$ for request set $\mathcal{R}$, time $t$, and batch size $B$), but such methods scale poorly with large request sets and cannot flexibly incorporate practical factors like preemptions.  

To address this, \algname uses a lightweight design that prioritizes requests by their \emph{margin goodput per unit bandwidth}:  
\[
Priority(r) = \frac{goodput_\Delta(r)}{bw_\Delta(r)} 
= \frac{goodput(r)}{t_{\text{gen}}(r)}.
\]
This formulation naturally prefers requests with high payoff relative to their bandwidth demand, while eliminating sensitivity to $\Delta$. To avoid starvation, including for best-effort requests without explicit SLOs, \algname inflates each deemed $goodput(r)$ by a small additive constant $\delta$ per frame, ensuring long-waiting requests eventually rise in priority. We later show that this heuristic achieves competitive performance guarantees in theory and near-optimal goodput empirically.

\paragraph{Grouped Margin Goodput Maximization for Batch Scheduling.}
Simply prioritizing requests by margin goodput can produce batches with highly heterogeneous input lengths, which degrades batching efficiency and ultimately reduces service goodput (Figure~\ref{fig:heter-homo}). Next, \algname extends individual prioritization into a \emph{grouped scheduling strategy} that jointly balances goodput payoff and length homogeneity.  

As illustrated in Figure~\ref{fig:bin-packing-sliding-window}, let $B$ denote the batch size. \algname first filters requests by retaining only those whose priority is at least $p \cdot Priority(r_{(B)})$, where $Priority(r_{(B)})$ is the $B$-th highest priority and $0 < p \leq 1$ is a tunable cutoff (e.g., 0.95). This ensures that subsequent group scheduling focuses on a promising candidate pool. The retained requests are then sorted by input length, and a sliding window of size $B$ traverses the list. For each candidate group $\mathcal{G} \subseteq \mathcal{R}$ of size $B$, \algname computes the aggregate priority:  
$
Priority(\mathcal{G}) = \sum_{r \in \mathcal{G}} Priority(r).
$
The group with the maximum $Priority(\mathcal{G})$ is selected as the execution batch, ensuring both high expected goodput and input-length alignment.  

The cutoff parameter $p$ controls the tradeoff: smaller $p$ admits more candidates, improving homogeneity at the expense of diluting group-level goodput with low-priority requests, while larger $p$ enforces stronger goodput guarantees but increases heterogeneity due to the long-tailed input length distribution (Figure~\ref{fig:heter-homo}). Fortunately, because LLM serving is long-running, \algname automates and continuously adapts $p$ online by exploring different thresholds and converging to those that maximize end-to-end goodput. 


\paragraph{Preemption to Correct Scheduling Errors.}
Optimal scheduling relies on both future arrivals and progressively refined request information, introducing uncertainty that can lead to suboptimal online decisions. Preempting running requests to correct these errors, however, incurs non-trivial overheads (e.g., batch stalls and KV cache swaps), which may outweigh benefits if not carefully controlled.

\name mitigates this with two safeguards. First, because preemption overhead scales with the KV cache size to be released, the resulting stall time is predictable given the KV cache size and I/O bandwidth. \name estimates the potential goodput loss as: 
$
goodput\_loss = stall\_duration \times token\_generation\_speed,
$
and performs preemption only when the projected gain from admitting a higher-priority request exceeds this cost, thereby ensuring a net benefit. Second, to prevent excessive churn, scheduling updates are restricted to discrete time frames (e.g., $\Delta = 50$ decoding steps; about 300 ms). This time-slicing aligns with the frame-based scheduling formulation, smooths execution, and allows any surplus bandwidth to be reclaimed in subsequent frames.  
Together, these mechanisms enable \name to correct scheduling decisions with negligible overhead (\(<1\%\)) in practice (\S\ref{eval:e2e}).

\paragraph{Achievable Guarantees.} 
We next analyze the scheduling efficiency and quality of \algname: (1) \emph{Scalability}: Given $N$ requests, computing the minimum serving bandwidth for each request requires $O(N)$ time, ordering their priorities adds $O(N \log N)$, and composing batches through sliding-length grouping incurs another $O(N)$. Overall, the scheduling process is bounded by $O(N \log N)$ complexity. Figure~\ref{fig:scheduling-scaling} shows that \algname scales efficiently, scheduling thousands of concurrent requests within 20 milliseconds, making it practical for online serving (\S\ref{eval:e2e}); (2) \emph{Quality}: Through the \emph{amortized analysis} method~\cite{amortized}, we prove  that \algname achieves a competitive performance guarantee relative to the even \emph{offline} optimal scheduler with future request arrival information. A detailed proof is provided in Appendix~\ref{app:JITServe}, specifically: 
\begin{theorem}
Let $G_{\text{\algname}}(\mathcal{R})$ denote the goodput achieved by online \algname on the request set $\mathcal{R}$, and $G^\star(\mathcal{R})$ corresponds to the goodput achieved by the optimal offline scheduler. Then we have a guarantee that 
$
G_{\text{\algname}}(\mathcal{R})
\;\ge\;
\frac{1}{8.56} \cdot G^\star(\mathcal{R}) \,.
$
\end{theorem}

\subsection{\name across Design Space}
\label{sec:design-dis}

An ideal scheduler must adapt to diverse LLM deployment scenarios without requiring reinvention, handling multiple models, ensuring fairness, and maintaining robustness under unfavorable SLO settings. 


\paragraph{Supporting Multiple Models.} 
Practical deployments often replicate models to scale throughput, with replicas potentially operating at different speeds (\eg, due to heterogeneous hardware or batch sizes). While \name scales efficiently---its request estimation and refinement can run in parallel across requests, and \algname achieves low computational complexity ($O(N\log N)$)---supporting multiple models introduces a new challenge: a single request $r$ may have different serving bandwidth requirements across model replicas due to varying generation speeds or data locality (\eg, KV cache). 

To address this, we extend \algname using a \emph{power-of-$K$} approach. For each request $r$, we create $K$ dummy copies $[r_1, \dots, r_K]$, by randomly sampling $K$ models from the $M$ available models. Each dummy $r$  carries a replica-specific priority $priority(r)$, and scheduling proceeds as usual over the enlarged set. Once a request is assigned to a replica, its other dummies are removed from the queue, incurring negligible overhead since no real LLM execution is involved. Thanks to \algname's strong scaling capability, $K$ can be set equal to $M$, ensuring full replica coverage. This multi-model extension increases scheduling complexity at most by $O(K)$ while aligning requests with their most favorable replicas, preserving provable performance guarantees. Empirical results confirm that \name consistently achieves superior performance in multi-model deployments (\S\ref{eval:ablation}).

\paragraph{Extending to Other Objectives.} 
Prioritizing requests solely by goodput can lead to unfairness 
and be vulnerable to outliers in unfavorable settings. For example, corrupted users may continuously submit requests with extremely
strict SLO demands to monopolize serving bandwidth, which again boils down to ensuring fairness in the wild. 

\name can seamlessly incorporate additional objectives, such as fairness, with minimal changes.  
Given a developer-specified fairness function $\text{Fair}(r)$, we redefine the request priority as
$
priority'(r) = (1-f) \cdot priority(r) + f \cdot \text{Fair}(r),
$
where $priority(r)$ is the default goodput density (\S\ref{sec:slo-scheduler}) and $f \in [0,1]$ balances efficiency and fairness. As $f \to 1$, requests with lower fairness attainment gain higher priority. This lightweight adaptation allows \name to enforce diverse fairness policies while offering flexible tradeoffs.


\section{Implementation}
\label{sec:implementation}

We implemented \name atop vLLM~\cite{vllm-git} with about 2,800 lines of code, preserving its  APIs for broad compatibility.

\paragraph{Execution Backend.}
\name augments the vLLM core engine with a policy module that generalizes the scheduler layer to support multiple scheduling policies, while preserving the efficiency guarantees of chunked-prefill execution. \name further inherits vLLM's prefix caching~\cite{sglang-neurips24} and sharing mechanisms to maximize reuse across overlapping requests. 
The module maintains a compact priority cache to amortize priority computations, updating only upon request arrivals or preemption events to reduce redundant overhead. 


\paragraph{Control Plane.}
The QRF-based length predictor and pattern graph matcher are offloaded to a separate asynchronous process through gRPC communication. The exchanged metadata is only a few bytes per event, making the communication overhead negligible. For robustness, it integrates a monitoring daemon that tracks component liveness and persists periodic metadata checkpoints, ensuring rapid state reconstruction and minimal recovery latency under failures. 
\name extends the OpenAI API~\cite{openai-pyapi} with SLO-aware parameters, specifically   
\textit{client.responses.create(model, input, deadline=None, target\_tbt=0.2, target\_ttft=5, waiting\_time=5)}.
For admission control, \name enforces a maximum waiting\_time (e.g., 5 seconds): requests unscheduled beyond it are dropped to prevent overload, ensuring predictable service behavior.

\section{Evaluation}
\label{sec:eval}

We evaluate \name with a variety of popular models and LLM applications.
Our main observations include:

\begin{denseitemize}
\item \name improves service goodput by 1.4$\times$-6.3$\times$ over existing advances, alternatively achieving 28.5\%--83.2\% resource savings to sustain the same goodput, while achieving near-oracle performance (\S\ref{eval:e2e}); 

\item \name effectively balances performance across diverse request types, achieving strong P50 metrics and comparable P95 tail latency for all request patterns (\S\ref{eval:break-down});

\item \name demonstrates robustness across different SLO requirements,  workload compositions, and distributed settings, consistently outperforming baselines  (\S\ref{eval:ablation}).
\end{denseitemize}

\subsection{Experimental Setup}
\label{sec:exp-setup}

We evaluate \name on a set of widely used LLMs with diverse architectures, including Llama-3.1-8B~\cite{llama_3}, Qwen2.5-14B~\cite{qwen2.5}, Qwen3-30B-MoE-A3B~\cite{qwen3}, and Llama-3.1-70B~\cite{llama_3}. These models span both dense and MoE designs as well as different parameter scales. Experiments are conducted on a cluster of 16 NVIDIA A100 GPUs.

\begin{table}
\centering
\small
\resizebox{0.99\linewidth}{!}{
\begin{tabular}{c|c|ccccc}
\toprule
\textbf{Workload} & \textbf{Req Type} & \textbf{Metric} & \textbf{Mean} & \textbf{Std.} & \textbf{P50} & \textbf{P95} \\
\midrule
\multirow{4}{*}{Chatbot}
& \multirow{2}{*}{Single} & Input & 93 & 244 & 27 & 391 \\
& & Output & 318 & 313 & 225 & 1024 \\
\cmidrule{2-7}
& \multirow{2}{*}{Compound} & Input & 1300 & 912 & 1097 & 2767 \\
& & Output & 4458 & 1176 & 4417 & 6452 \\
\midrule
\multirow{4}{*}{\begin{tabular}[c]{@{}c@{}}Deep\\Research\end{tabular}} 
& \multirow{2}{*}{Single} & Input & 1911 & 2781 & 403 & 7573 \\
& & Output & 534 & 644 & 410 & 1544 \\
\cmidrule{2-7}
& \multirow{2}{*}{Compound} & Input & 12223 & 8407 & 10807 & 29282 \\
& & Output & 3541 & 2370 & 3148 & 7525 \\
\bottomrule
\end{tabular}
}
\caption{Our evaluations include four popular applications: Chatbot, Deep Research, Agentic CodeGen, and Math Reasoning. This table shows example request length statistics for two of them. }
\label{tab:dataset_stats}
\end{table}

\paragraph{Workloads.}
To construct the three request patterns, we use the Alpaca~\cite{alpaca-23} and LMsys-chat~\cite{lmsys-chat-23} datasets to build the Chatbot application. We further incorporate a long-context math reasoning application~\cite{tot-neurips-23}, a deep research application based on the Search Arena benchmark~\cite{searcharena2025}, and an agentic code generation application~\cite{autogen-arxiv23}.  
From LMsys-chat usage analysis~\cite{lmsys-chat-23}, we extract the distribution of real-world use cases. Requests are tagged according to statistics from our user study (Table~\ref{tab:user_survey}). For example, 38.1\% of code generation requests are classified as latency-sensitive, while the deep research application is modeled as compound requests. Table~\ref{tab:dataset_stats} reports the request characteristics for two of all four applications.
Request arrivals follow Microsoft's real-world LLM serving trace, scaled to match our cluster resources. Following prior advances~\cite{vllm-sosp23,llumnix-osdi24}, We also perform ablation studies with arrivals generated using a Poisson distribution. Each evaluation run involves more than 10K requests over an online deployment window of at least one hour.  

We set request SLOs using the P95 latencies measured from 1K DeepSeek API calls. This results in latency-sensitive requests requiring $\sim$2s TTFT and $\sim$100ms TBT, while deadline-sensitive requests have an E2EL of 20s. For compound requests, the E2EL SLO is scaled with the number of stages, defined as $20 \times$ (number of stages) seconds.  

Unless otherwise noted, we adopt a 1:1:1 ratio across the three request patterns, which yields a workload mix dominated by latency-sensitive requests. We also show that \name achieves consistent improvement across different settings of SLO requirements and workload compositions (\S\ref{eval:ablation}).

\begin{figure*}[!t]
    \centering
    \includegraphics[width=\linewidth]{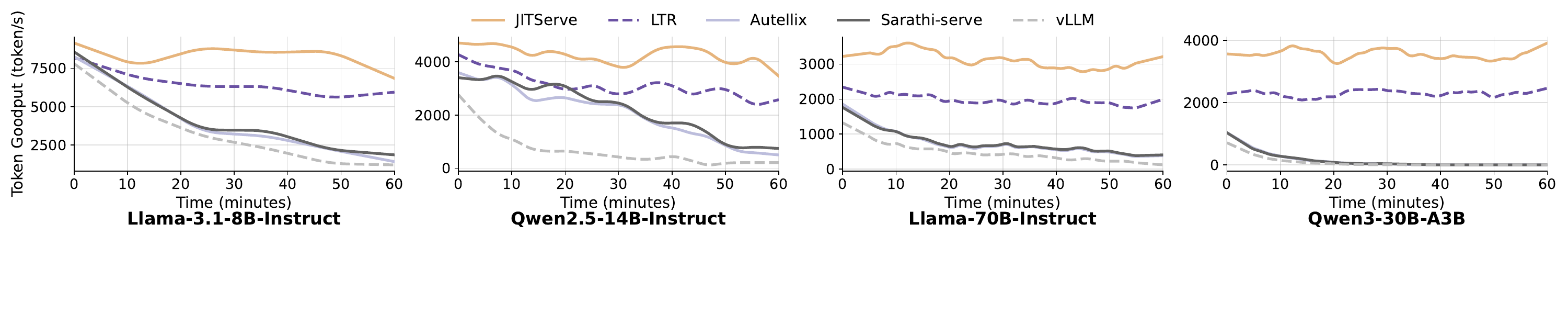}
    \vspace{-1.1cm}
    \caption{Service goodput over time in a one-hour online serving experiment. \name achieves consistently high service good while the baselines suffer cascading SLO violations and degraded service goodput over time.}
    \label{fig:e2e-timeline}
\end{figure*}

\begin{figure}
    \includegraphics[width=\linewidth]
    {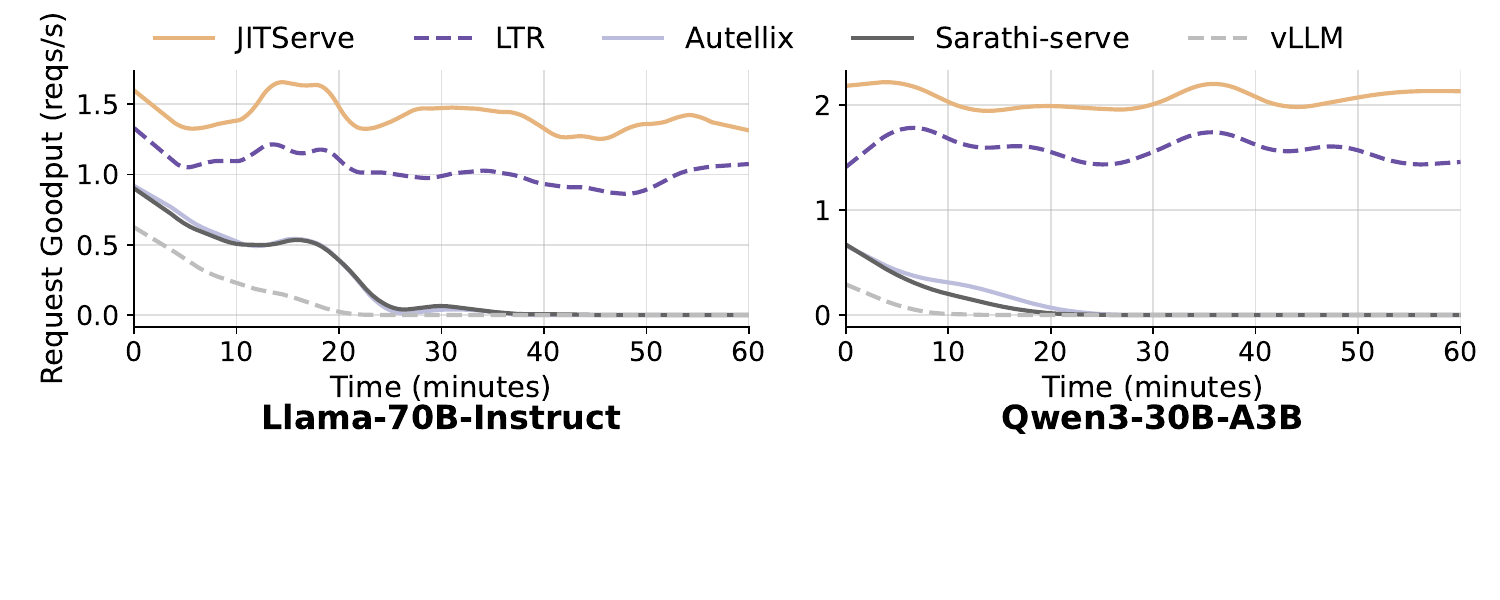}
    \vspace{-1.1cm}
    \caption{\name achieves consistently better request-level SLO service goodput in online deployments.}
    \label{fig:e2e-goodput}
\end{figure}

\paragraph{Baselines.} 
We evaluate against four state-of-the-art designs:
\begin{denseitemize}
    \item \emph{Autellix}~\cite{autellix-arxiv25}: Uses Program-level Least Attained Service scheduling (PLAS) to imitate SJF, optimizing request completion time (E2EL) for compound workloads.
    \item \emph{Learn to Rank (LTR)}~\cite{ltr-nips24}: Leverages an LLM prediction model to predict the relative response length ranking of requests and prioritize the smallest one, imitating SJF.
    \item \emph{vLLM}~\cite{vllm-sosp23}: A recent advanced LLM serving backend with continuous batching~\cite{orca-osdi22} and PagedAttention~\cite{vllm-sosp23}. Uses FCFS scheduling.
    \item \emph{Sarathi-Serve}~\cite{sarathi-serve-osdi24}: Extends vLLM with chunked prefills to optimize TTFTs and TBTs for requests.

\end{denseitemize}

\paragraph{Metrics.}
We focus on higher \emph{service goodput}. Since our design is agnostic to the definition of goodput (\S\ref{sec:overview}), we choose two popular goodput definitions focus on different levels: (1) \emph{Token-level goodput}: the number of tokens meeting SLO requirements, following popular SLO preferences; and (2) \emph{Request-level goodput}: the number of requests meeting the SLO requirements. We also report traditional metrics such as TTFT, TBT, and throughput for performance breakdown.

All results are averaged over five independent runs.

\subsection{End-to-End Performance}
\label{eval:e2e}
We start with end-to-end evaluations in online deployments. 

\paragraph{\name substantially improves service goodput.}
We first deploy \name in a one-hour online experiment to evaluate its long-term performance. As shown in Figure~\ref{fig:e2e-timeline}, \name consistently achieves high service goodput over time, outperforming LTR by 1.3$\times$--1.7$\times$ and Autellix by 5.3$\times$--6.1$\times$. This improvement arises from \name's ability to dynamically prioritize requests based on per-request SLOs and their serving bandwidth requirements.

In contrast, existing systems such as Sarathi-Serve and vLLM suffer from increasing head-of-line (HOL) blocking, leading to cascading SLO violations and reduced service goodput over time. \name mitigates this degradation by leveraging conservative upper-bound length predictions and maximizing residual bandwidth for other requests, resulting in stable service goodput throughout the evaluation. Notably, \name maintains high service goodput consistently over the entire one-hour deployment.

In addition to token-level goodput improvement, we also measure request-level goodput. As shown in Figure~\ref{fig:e2e-goodput}, \name achieves 2.3$\times$-4.5$\times$ higher goodput than LTR.

\begin{figure}
  \centering
  \begin{minipage}{0.48\linewidth}
   \centering
   \includegraphics[width=\linewidth]{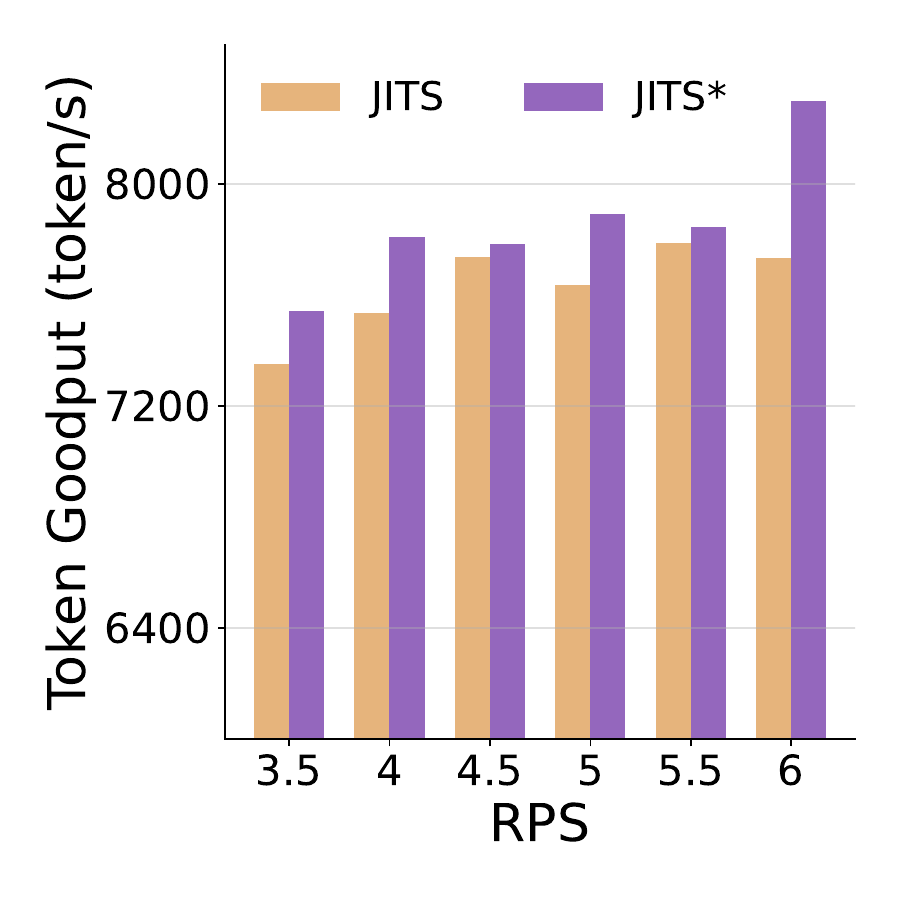}
   \vspace{-.6cm}
   \captionof{figure}{\name achieves close-to-oracle performance.}
  \label{fig:e2e-oracle}
  \end{minipage}
  \hfill
  \begin{minipage}{0.48\linewidth}
\centering
\includegraphics[width=\linewidth]{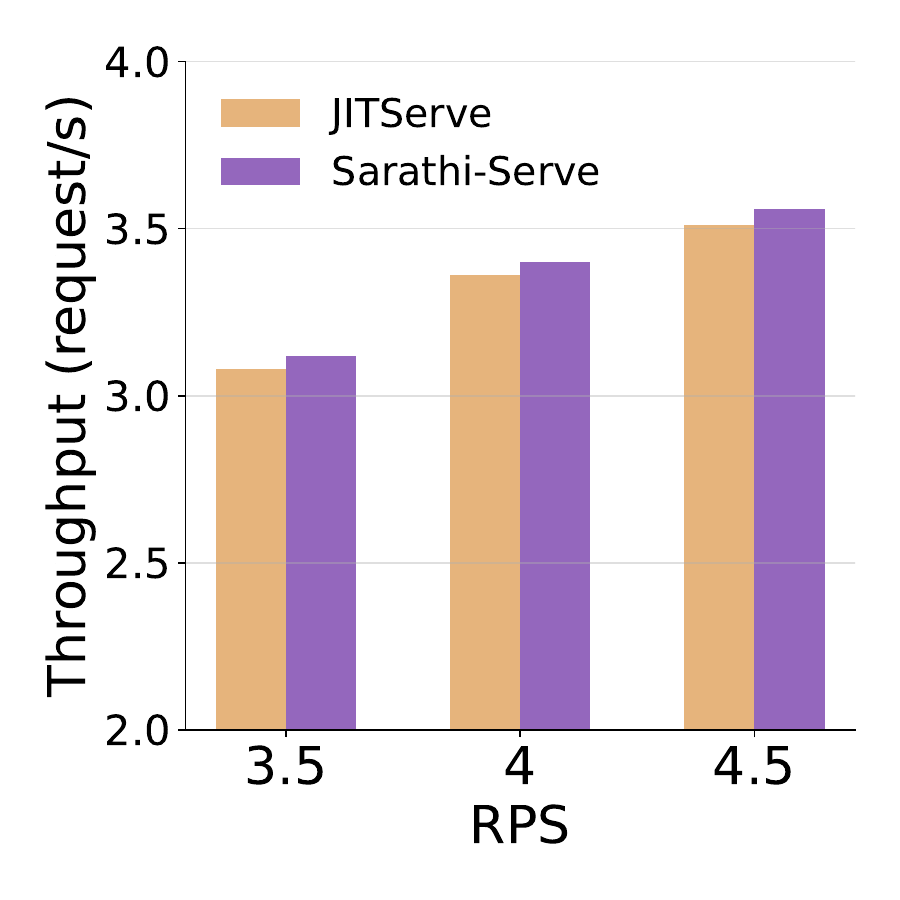}
\vspace{-.6cm}
   \captionof{figure}{\name introduces little overhead in throughput.}
   \label{fig:e2e-throughput}
   \end{minipage}
\end{figure}

\paragraph{\name achieves near-oracle performance.}
We further benchmark \name against an oracle variant, \name*, which operates with perfect foresight of request information (\ie, response length and execution graph) across varying request-per-second (RPS) settings. As shown in Figure~\ref{fig:e2e-oracle}, \name achieves performance within 3--9\% of the oracle despite relying on imperfect predictions. This robustness arises from two key factors: (1) the Request Analyzer generalizes effectively to unseen workloads and request mixes, and (2) the scheduling policy is designed to tolerate uncertainty, leveraging approximate information and progressively relaxing the conservatism to make near-optimal decisions. Together, these results demonstrate that \name approaches the best achievable performance without strong assumptions.

\paragraph{\name does not hurt system throughput.}
A common concern with sophisticated scheduling is the potential throughput loss. To evaluate this, we compare \name against Sarathi-Serve, which employs FIFO scheduling without preemption and thus represents a near upper-bound on serving throughput.
As shown in Figure~\ref{fig:e2e-throughput}, \name achieves comparable throughput to Sarathi-Serve across different request-per-second (RPS) settings, reaching 96\%--98\% of its performance. This demonstrates that \name's additional modules incur negligible overhead, aided by its cost-aware design that selectively corrects scheduling errors only when the potential goodput gains outweigh preemption costs (\S\ref{sec:slo-scheduler}).

\paragraph{\name sustains high goodput under load surges.}
\begin{figure}
    \centering
    \includegraphics[width=\linewidth]{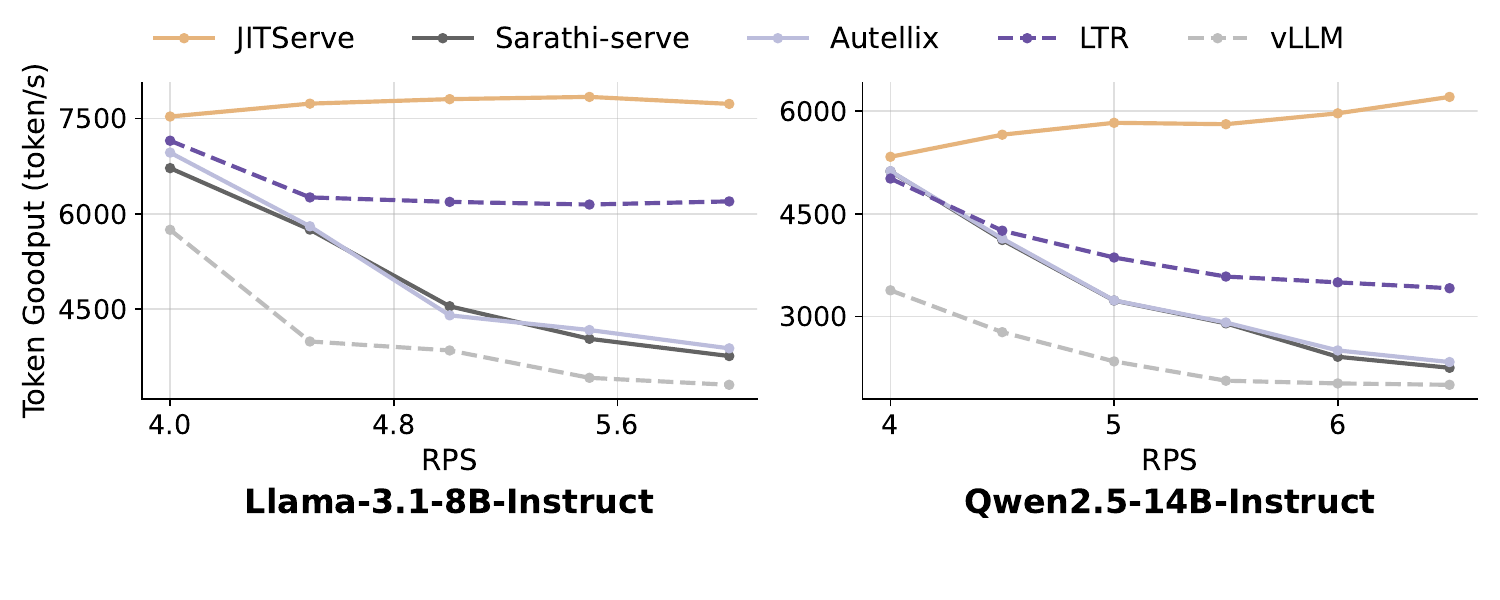}
    \vspace{-.8cm}
    \caption{\name sustains high goodput across request loads.}
    \label{fig:ablation-rps}
\end{figure}
To evaluate scalability and robustness, we measure service goodput under varying request arrival rates. As shown in Figure~\ref{fig:ablation-rps}, all baselines exhibit sharp performance drop as system load increases due to contention. In contrast, \name consistently achieves the highest  goodput across all load levels by dynamically adjusting priorities and resource allocations. 

As a result, \name significantly mitigates the impact of increasing request rates on goodput degradation, confirming its suitability under real-world load dynamics.

\subsection{Performance Breakdown}
\label{eval:break-down}
We next analyze \name's performance by breaking it down into two key aspects: its ability to handle different request types, and its performance when key components are ablated. 
By understanding these aspects, we gain deeper insights into \name's effectiveness and how it performs in various scenarios. 

\begin{figure}
    
    \subfigure[Latency-sensitive TTFT (s)\label{fig:latency-ttft}]{\includegraphics[width=0.48\linewidth]{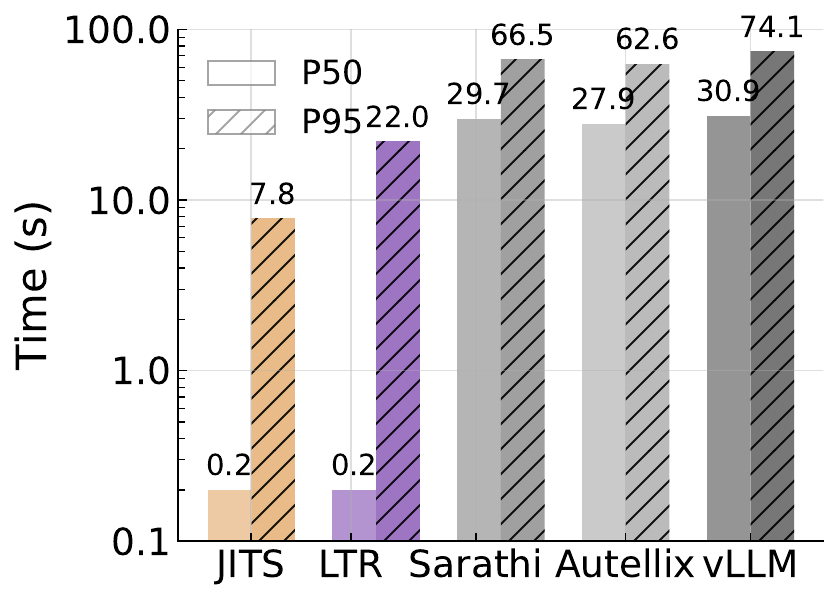}}
    \hfill
    \subfigure[Latency-sensitive TBT (ms)\label{fig:latency-tbt}]{\includegraphics[width=0.48\linewidth]{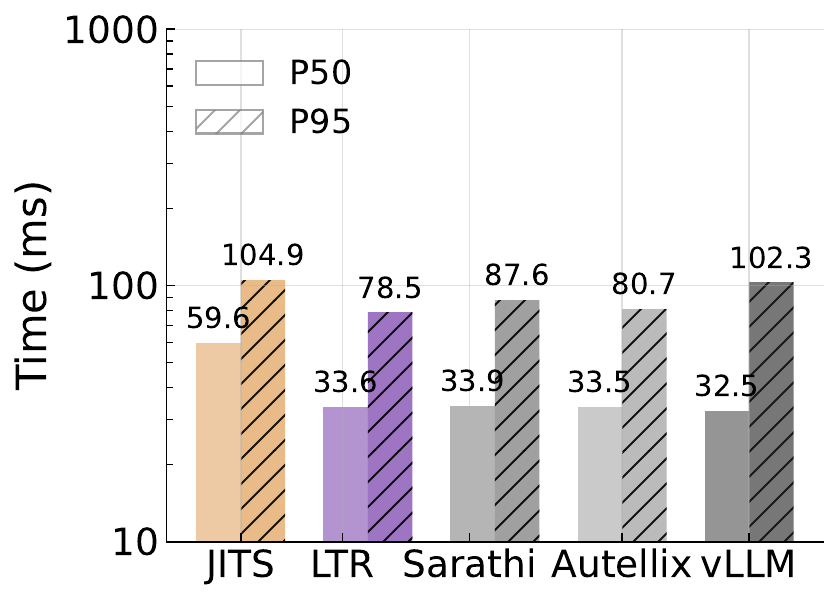}}
    
    \subfigure[Deadline-sensitive E2EL (s)\label{fig:throughput-ttlt}]{\includegraphics[width=0.48\linewidth]{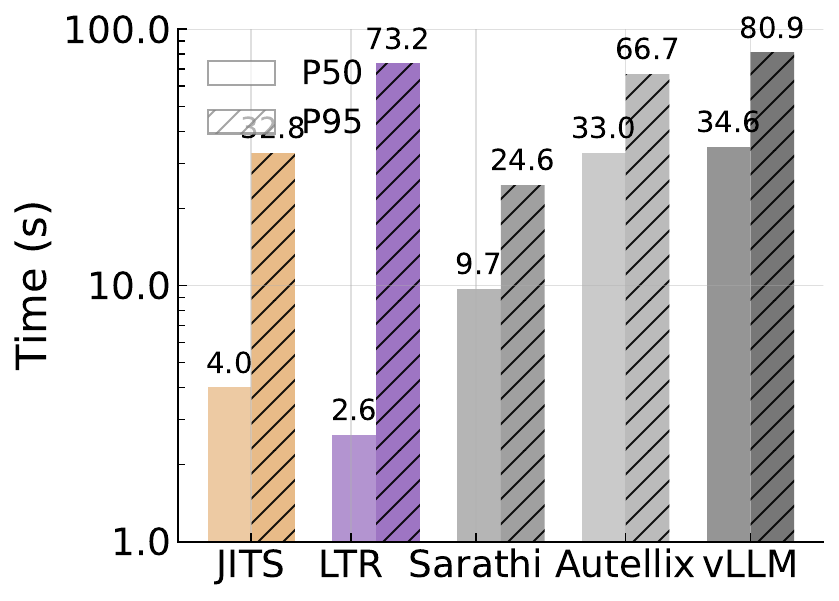}}
    \hfill
    \subfigure[Compound E2EL (s)\label{fig:collective-ttlt}]{\includegraphics[width=0.48\linewidth]{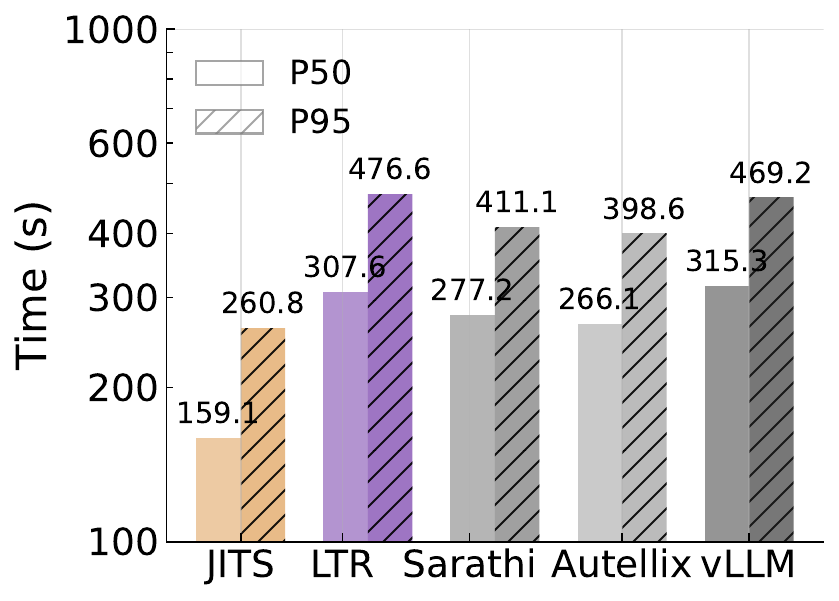}}
    
    \caption{
        Performance metrics comparison across different baselines.
        \subref{fig:latency-ttft} Time to First Token, 
        \subref{fig:latency-tbt} Time Between Tokens,
        \subref{fig:throughput-ttlt} Deadline-Sensitive Request Latency,
        \subref{fig:collective-ttlt} Compound Request Latency.
        All metrics are shown in log scale with P50/P95 percentiles.
    }
    \label{fig:breakdown-metrics}
\end{figure}


\paragraph{Breakdown by Request Types.} 
While we have demonstrated \name's large goodput improvement, we next study how \name handles diverse SLOs across different request types, using conventional performance metrics such as TTFT, TBT, and E2EL. 
As shown in Figure~\ref{fig:latency-ttft}, \name excels at minimizing TTFT for latency-sensitive requests, demonstrating its ability to deliver responsive service under strict SLOs. Notably, this is achieved without incurring excessive TBT (Figure~\ref{fig:latency-tbt}). For deadline-sensitive and compound requests, \name achieves favorable median E2EL (Figures~\ref{fig:throughput-ttlt} and~\ref{fig:collective-ttlt}), ensuring that many requests meet their SLOs without oversubscribing bandwidth merely to minimize the average E2EL. Across all request types, \name maintains strong P95 performance, highlighting the effectiveness of its conservative scheduling while avoiding starvation.

Another common concern in LLM scheduling is whether optimization comes at the expense of a small subset of requests. Our results show that \name avoids this pitfall: it significantly reduces tail latency (P95) compared to baselines like SJF and Autellix, indicating that it does not sacrifice much fairness for efficiency. By integrating deadline-awareness and service gain estimation, \name mitigates starvation and ensures that SLO-critical requests are not blocked by long-running tasks.
In contrast, LTR shows very competitive E2EL on deadline-sensitive requests (Figure~\ref{fig:limitation_existing_solutions}) as by design it prioritizes the request with potentially shortest response, thus achieving strong average E2EL performance. However, it struggles under diverse workloads: it oversubscribes bandwidth to deadline-sensitive requests which degrades overall goodput.

\paragraph{Breakdown by Components.}
\begin{figure}
    \centering
    \includegraphics[width=\linewidth]{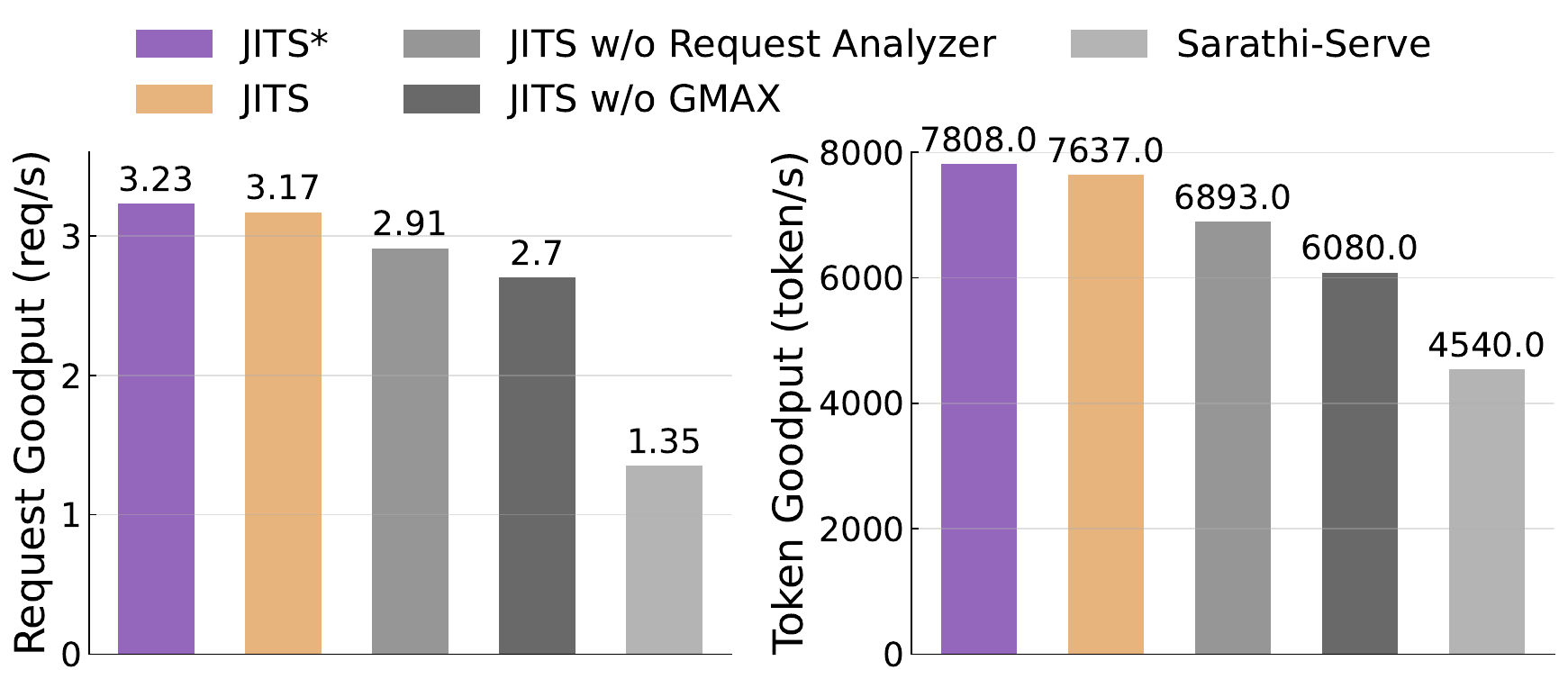}
    \caption{Request analyzer and GMAX algorithm jointly contribute to \name's high-quality, resilient serving.}
    \label{fig:breakdown-component}
\end{figure}

Figure~\ref{fig:breakdown-component} presents a component-level breakdown of \name and several ablated variants: (1) \name with precise knowledge (JITServe*), (2) \name without the Request Analyzer (falling back to average response length estimation), (3) \name without \algname scheduling (replaced by SJF scheduling based on Request Analyzer estimates), and Sarathi-Serve.  
Beyond the near-oracle performance achieved by \name in terms of SLO goodput, we observe that removing either the Request Analyzer or \algname results in a noticeable degradation in goodput, highlighting their essential roles in the design.



\subsection{Sensitivity and Ablation Studies}
\label{eval:ablation}

%

\begin{figure}
    \centering
    \includegraphics[width=\linewidth]{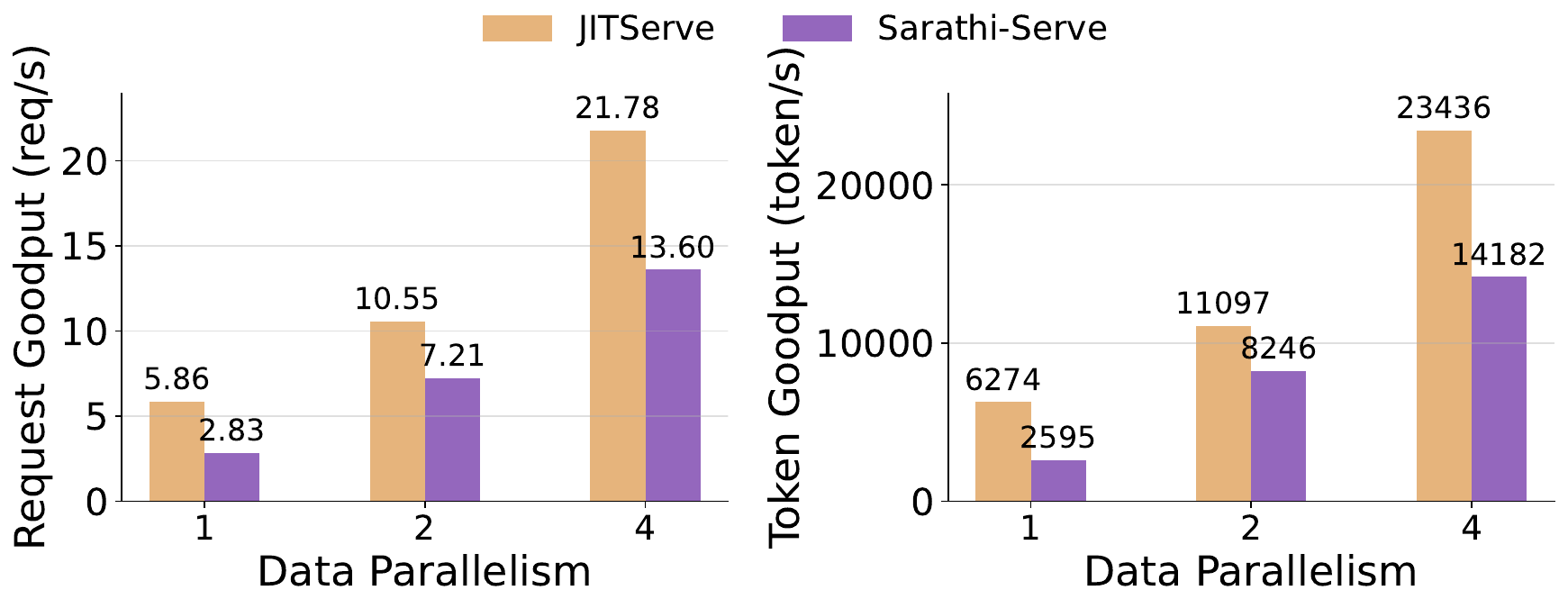}
    \caption{\name scales effectively to multi-model deployments.}
    \label{fig:multi-model}
\end{figure}

\paragraph{Extending to Multiple Models.}
We next investigate \name's performance when serving multiple model replicas with data parallelism. We scale the request arrival rates proportionally to the number of model replicas. As shown in Figure~\ref{fig:multi-model}, while all systems achieve higher goodput with additional replicas, \name consistently outperforms the baseline by 1.34$\times$--2.42$\times$ across all configurations.

\begin{figure}[t]
    \centering
    \includegraphics[width=\linewidth]{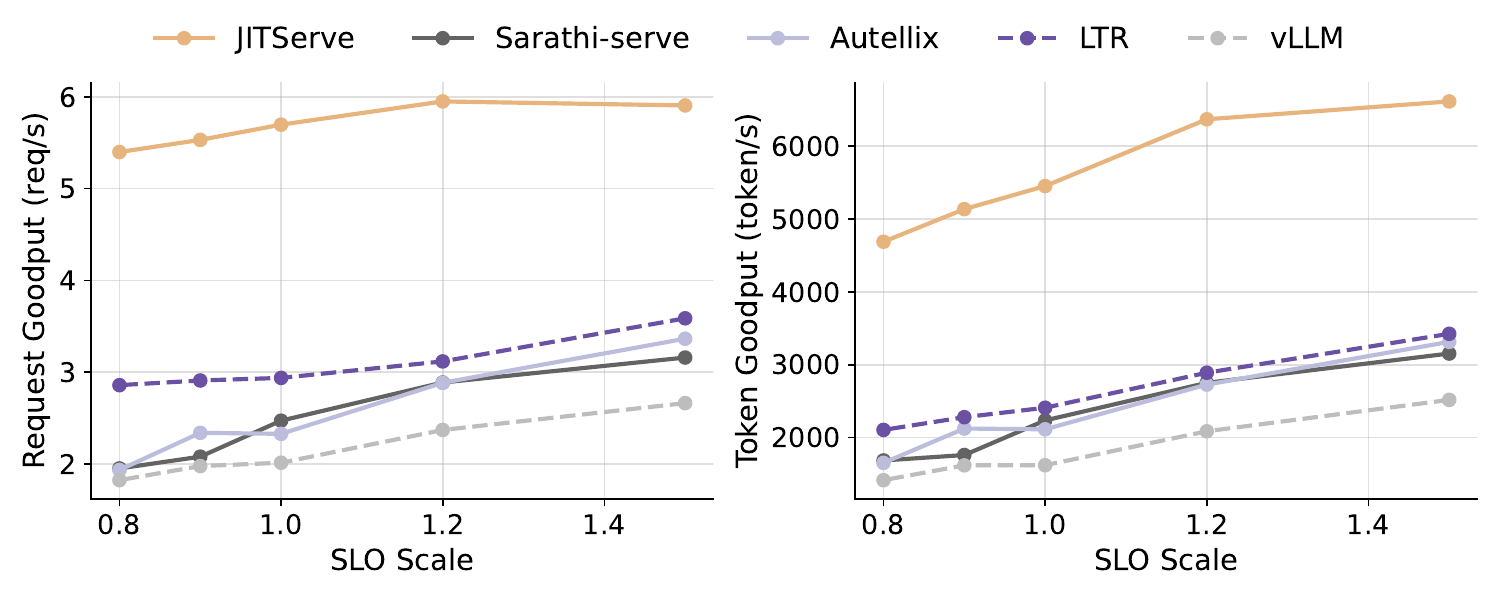}
    \caption{\name outperforms across various SLO tightness.}
    \label{fig:slo-impact}
\end{figure}

\paragraph{Impact of SLO Constraints.} 
We evaluate how \name responds when the SLO requirements are uniformly relaxed across all request types. The SLO constraints are scaled by a common factor (e.g., $0.8\times, 1.5\times$), as users or applications may tolerate varying response times. 
As shown in Figure~\ref{fig:slo-impact}, relaxing SLO constraints naturally improves the SLO goodput. \name consistently improves both request and token goodput by 2.3$\times$--2.8$\times$ over existing advances. 

\paragraph{Impact of Workload Composition.}

\begin{figure}[t]
  \centering
  \begin{minipage}{0.5\linewidth}
\centering
\includegraphics[width=\linewidth] {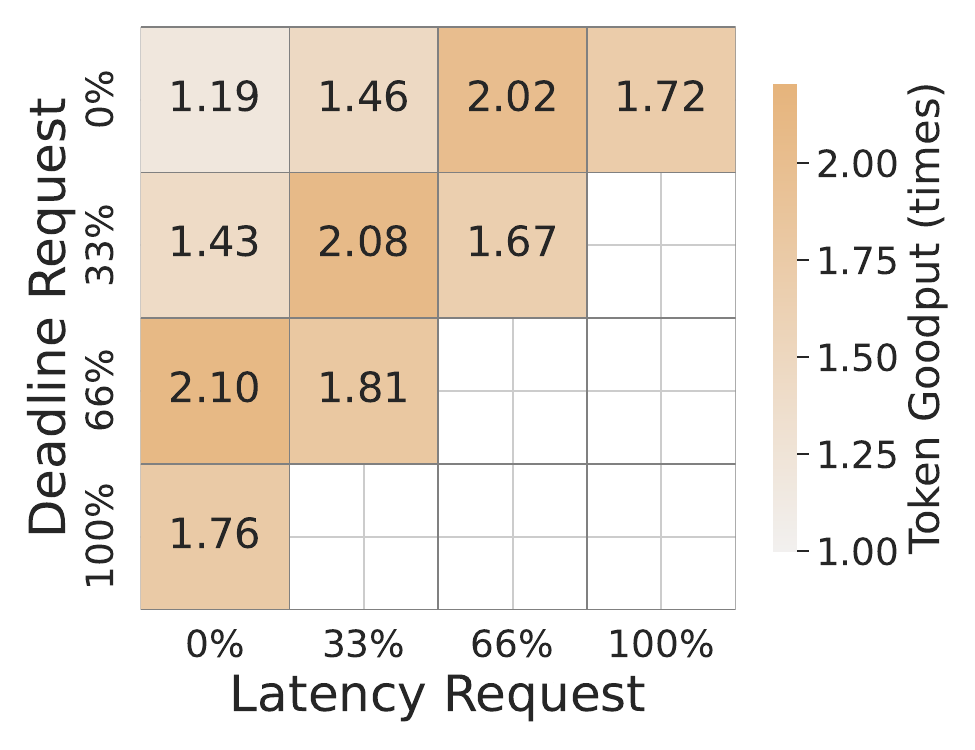}
   \captionof{figure}{\name maintains higher performance across varying workload compositions.}
   \vspace{-0.1em}
   \label{fig:ablation-workload}
   \end{minipage}
   \hfill 
   \begin{minipage}{0.48\linewidth}
   \centering
\includegraphics[width=\linewidth]{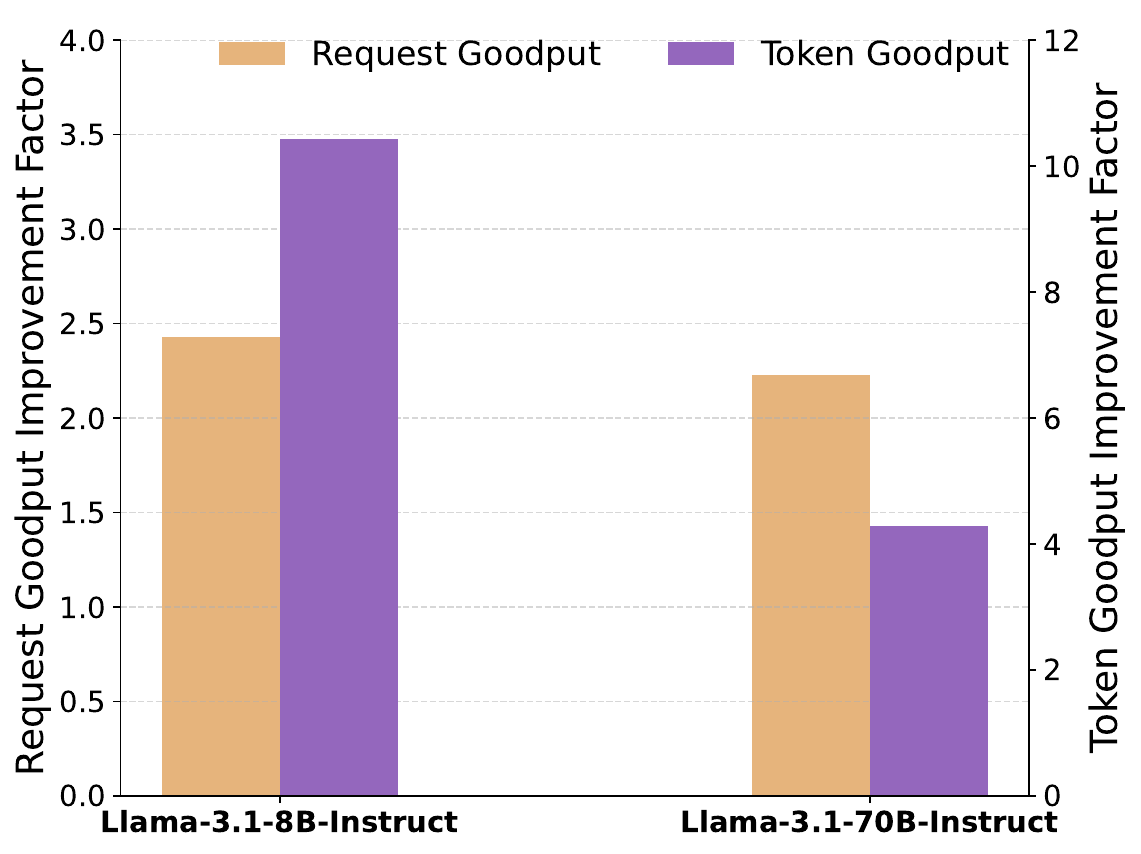}
    \caption{\name outperforms earliest deadline first.}
    \label{fig:edf-impact}
   \vspace{0.3em}
  \label{fig:ablation-burstgpt}
  \end{minipage}
\end{figure}

We next study different workload compositions, under a wide range of workload mixes, including settings dominated by a single request type (e.g., 0\% latency-sensitive workloads) and heterogeneous mixtures. Across all cases, Figure~\ref{fig:ablation-workload} shows that \name consistently achieves higher service goodput than existing approaches, such as  1.8$\times$ goodput improvement in a setting with 33\% latency- and 66\% deadline-sensitive workloads. Notably, \name outperforms Sarathi-Serve by $1.72\times$ even for latency-sensitive-only requests, its intended design point. 



\paragraph{Comparison to Earliest-Deadline First.} 
We here summarize the performance comparison to the earlies-deadline first (EDF) policy. Figure~\ref{fig:edf-impact} shows that \name achieves better goodput since EDF does not account for the heterogeneous SLOs and goodput. 


\section{Related Work}
\label{sec:related}
\paragraph{LLM Serving System.} 
Recent advancements in LLM have led to the development of numerous inference systems.
Orca~\cite{orca-osdi22}, dLoRA~\cite{dlora-osdi24}, VTC~\cite{VTC-osdi24}, and FastServe~\cite{FastServe-arxiv} present varied strategies for batching, concurrent serving, and fairness.
vLLM~\cite{vllm-sosp23} and InfiniGen~\cite{infinigen-osdi24} focus on KV-cache management to improve memory utilization and throughput.
DistServe~\cite{distserve-osdi24}, Sarathi-Serve~\cite{sarathi-serve-osdi24}, Llumnix~\cite{llumnix-osdi24}, and Splitwise~\cite{splitwise-isca24} capitalize on characteristics of the prefilling and decoding stages to minimize TTFT and TBT latency.
Some other systems explore GPU kernel optimizations~\cite{fastertransformer, turbotransformers-ppopp21,flashattention-22,flashattention-23,flashattention-24}, model parallelism~\cite{alpaserve-osdi23, deepspeed-sc22}, and preemptive scheduling~\cite{spotserve-asplos24}.
However, most prior works primarily target single-request latency without accounting for the diverse requirements of different applications. Parrot~\cite{parrot-osdi24} leverages the interconnections within LLM applications. Concord builds on these approaches, categorizing LLM applications into three types and addressing their SLO requirements collectively.
Parrot~\cite{parrot-osdi24} offers APIs to extract LLM request execution dependency. 
\name builds on them, categorizing LLM applications and addressing their SLO requirements collectively.

\paragraph{LLM Output Length Prediction.} 
Recent efforts such as TetriInfer~\cite{tetriinfer-arxiv24}, $S^3$\cite{s3-neurips23}, and u-Serve\cite{bert-predict-arxiv24} have proposed training multi-class classifiers (e.g., based on BERT) to predict LLM output length. However, these approaches are resource-intensive, both in training and during online inference. Moreover, our analysis reveals that existing classifiers exhibit significant deviations in prediction accuracy, ultimately leading to suboptimal scheduling decisions (\S\ref{sec:background}).
\name adopts a lightweight QRF model to estimate an upper bound on the output length, enabling conservative yet adaptive scheduling.

\paragraph{SLO-aware Resource Scheduling.} 
Satisfying SLO requirements has long been a central challenge in resource scheduling. 
In networking, 
Karuna~\cite{Karuna} performs deadline-aware flow scheduling to reserve bandwidth for best-effort flows, while QCLIMB~\cite{qclimb-nsdi24} uses QRF models to predict lower bounds on flow sizes for flow scheduling.
CASSINI~\cite{cassini-nsdi24} schedules ML training traffic across jobs to reduce network contention, and 
Caladan~\cite{caladan-osdi20} mitigates resource contention in hyperthreads to reduce OS tail latency.
AdaServe~\cite{adaserve-arxiv25} supports customizable SLOs through fine-grained speculative decoding. \name complements AdaServe by orchestrating a broader range of diverse SLO requirements across different LLM request types, aiming to maximize service goodput without request information.


\section{Conclusion}
\label{sec:outtro}

We introduce \name, an LLM request scheduler designed to maximize service goodput. \name conservatively estimates request characteristics and incrementally refines these estimates. It employs a novel \emph{grouped margin goodput maximization} algorithm that determines each request's minimum serving bandwidth needed while prioritizing requests with high grouped goodput payoff relative to their bandwidth when forming batches. Our evaluations across a variety of LLM applications and models demonstrate substantial improvements across a wide range of LLMs and applications.

\label{EndOfPaper}

{
\bibliographystyle{plain}
\bibliography{main}
}

\clearpage

\appendix

\section{Pattern-Graph Matching Alternatives}
\label{app:pattern-graph}

\begin{figure}[h]
  \centering
  {
    \subfigure[Illustration of $\phi(s)$. \label{fig:stage-ratio-explain}]{
      \IfFileExists{Figures/design/stage_ratio_explain.pdf}{
        \includegraphics[trim=120 260 380 150,clip, width=0.45\linewidth]{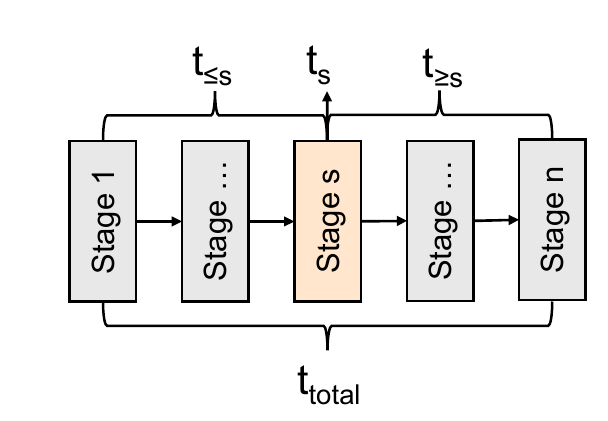}
      }{
        \fbox{\parbox[c][1.8in][c]{0.49\linewidth}{\centering Missing figure: stage\_ratio\_explain.pdf}}
      }
    }
    \subfigure[Estimation accuracy of different design. \label{fig:match-accuracy-three-design}]{
      \IfFileExists{Figures/design/deepresearch_online_matching_ddl_modes.pdf}{
        \includegraphics[width=0.45\linewidth]{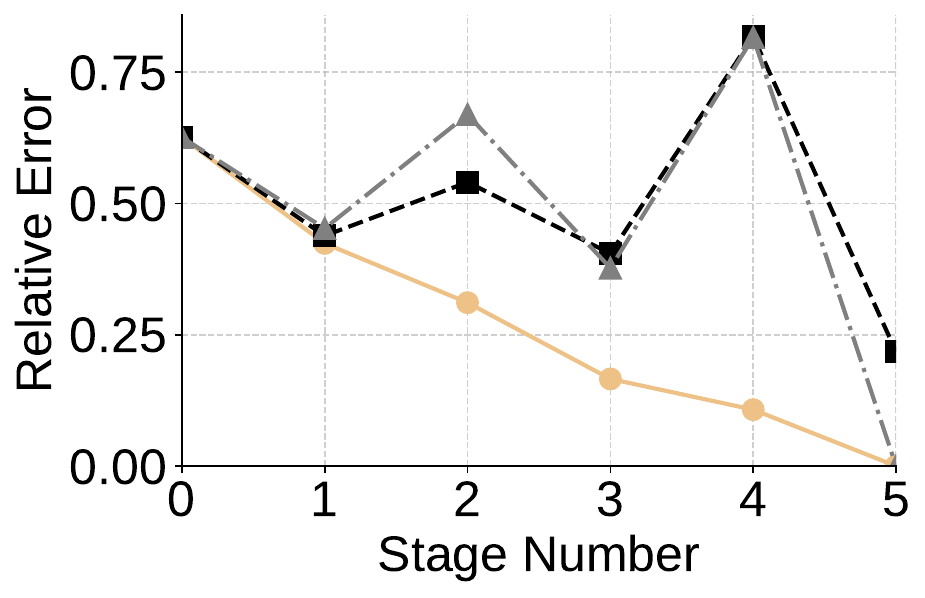}
      }{
        \fbox{\parbox[c][1.8in][c]{0.49\linewidth}{\centering Missing figure: deepresearch\_online\_matching\_ddl\_modes.pdf}}
      }
    }
  }
  \caption{Illustration and impact of different sub-deadline formulations.}
  \label{fig:subdeadline-variants}
\end{figure}

For completeness, we also considered two alternative pattern-matching formulations (\S\ref{subsec:request-analyzer}): setting $D_s$ proportional to $t_s/t_{\text{total}}$, and setting $D_s$ proportional to $t_s / t_{\geq s}$, where $t_{\geq s}$ is the accumulated time from stage $s$ to the end. However, as shown in Figure~\ref{fig:match-accuracy-three-design}, which reports relative error under online graph matching using traces from deepresearch requests, our design (orange curve) achieves substantially higher estimation accuracy than these alternatives.

\section{Symbols and Notations}
We will use the following notations for each request \(k\):
\emph{input length} \(L_i(k)\),
\emph{output length} \(L_o(k)\),
\emph{start time} \(s(k)\),
\emph{end time} \(e(k)\),
\emph{computing time} \(t_{\mathrm{comp}}(k)>0\),
\emph{service level objective~(SLO) time} \(t_\mathrm{SLO}(k)>0\),
\emph{remaining computing time} \(t^\mathrm{r}_\mathrm{comp}(k) > 0\),
\emph{remaining time to service level objective~(SLO)} \(t^\mathrm{r}_{\mathrm{SLO}}(k)>0\),
\emph{base goodput}
\begin{equation}
	R(k)
	\; \coloneqq \;
	\omega_i L_i(k) + \omega_o L_o(k)
\end{equation}
and \emph{scheduling indicator}
\begin{equation}
	I(k)
	\; \coloneqq \;
	\frac{R(k)}{t^\mathrm{r}_\mathrm{comp}(k) + \epsilon}
\end{equation}
together with \(\displaystyle t^\mathrm{r}_\mathrm{SLO} - t^\mathrm{r}_\mathrm{comp}(k) \ge 0\) scheduling filter and \(\epsilon > 0\) to avoid division by \(0\) error.
In our setting, a request will only realize its goodput if and only if it completes by its SLO; otherwise it will realize \(0\) goodput;
We $\sigma$ to denote a schedule, a sequence of served requests $(R, t_R)$, where $t_R$ refers to the served time of request set $R$ in schedule $\sigma$.
Note that since we allow preemption, some requests may get preempted and never complete; we use $\sigma^c$ to denote the set of completed requests in $\sigma$ and $\sigma^p \subset \sigma$ to denote the set of preempted requests.
For a request \(k\), denote by \(C_k\) its completion time (if it completes under \(\sigma\)), and by
\begin{equation}
	\mathsf{Goodput}(\sigma)\;\coloneqq\;\sum_{k:\, C_k\le t_{\mathrm{SLO}}(k)} R(k)
\end{equation}
the realized total on-time goodput of schedule \(\sigma\).

\section{Complexity Analysis}

\subsection{Optimal Scheduling}
\label{sec:OPT}
\begin{theorem}[NP-Hardness of optimal scheduling]
	Consider the following description of the previous LLM serving problem: given $n\in\mathbb{Z}_{>0}$ identical serving slots and a finite set of requests $R$, each request $r \in R$ specified by a computing time $t_{\mathrm{comp}}(k)>0$, a start time $s(k)$, an SLO time $t_{\mathrm{SLO}}(k)$, and a goodput $R(k)>0$ that is realized if and only if request $r$ completes by its SLO on time, decide a schedule $\sigma$ such that
	\begin{equation}
		\sigma = \arg \max_\sigma \; \mathsf{Goodput}(\sigma)
	\end{equation}
\end{theorem}

\begin{proof}
	It is well-known that the Multiple Knapsack Problem is NP-hard.
	We will reduce the Multiple Knapsack Problem to the LLM serving problem to show that the latter is NP-hard.
	\paragraph{Multiple Knapsack Problem}
	Given \( n \) knapsacks, each with a capacity \( \mathcal{C} \), and a set of items \( I \), each item \( i \in I \) has a size \( w_i > 0 \) and a value \( v_i > 0 \), and a target value \( \Psi \), determine whether there is a subset of items that can be assigned to the \( n \) knapsacks such that each knapsack's total size does not exceed \( \mathcal{C} \) and the total value is at least \( \Psi \).

	\paragraph{Reduction Construction}

	Given an instance of the Multiple Knapsack Problem, we construct an instance of the LLM serving problem as follows:

	\begin{itemize}[leftmargin=*]
		\item Set the number of serving slots \( n \) equal to the number of knapsacks.
		\item For each item \( i \in I \), create a request \( r_i \) with:
		      \[
			      t_{\mathrm{comp}}(r_i) = w_i, \quad s(r_i) = 0, \quad t_{\mathrm{SLO}}(r_i) = C, \quad R(r_i) = v_i.
		      \]
		\item Set the target total goodput to \( \Psi \).
	\end{itemize}

	\paragraph{Correctness of reduction~($\Rightarrow$)}
	If there is a solution to the Multiple Knapsack Problem,
	i.e., there exists a way to assign the items to the knapsacks such that each knapsack's total size does not exceed \( \mathcal{C} \) and the total value is at least \( \Psi \), then we can construct a valid schedule for the LLM serving problem.
	Each knapsack corresponds to a serving slot, and each item corresponds to a request.
	Since each knapsack has capacity \( \mathcal{C} \), and the total size of items assigned to each knapsack is at most \( \mathcal{C} \), the total processing time of requests in each slot does not exceed \( \mathcal{C} \), and all requests are completed on time.
	Therefore, the total goodput is at least \( \Psi \).
	\paragraph{Correctness of reduction~($\Leftarrow$)}
	Conversely, if there is a solution to the LLM serving problem such that the total goodput is at least \( \mathcal{C} \), then we can assign items to knapsacks such that each knapsack's total size does not exceed \( \mathcal{C} \), and the total value of the selected items is at least \( \Psi \).

	Thus, solving the LLM serving problem is equivalent to solving the Multiple Knapsack Problem.
	Since the Multiple Knapsack Problem is NP-hard, the LLM serving problem is also NP-hard.
\end{proof}

\section{Competitive Ratio Analysis}

\subsection{Analysis of Popular Scheduling Policies}
\label{app:prove-prior}
\subsubsection{Analysis of Earliest Deadline First Scheduling}
\label{sec:EDF}
\begin{theorem}[Non-competitiveness of EDF]
	The scheduling of Earliest Deadline First (EDF) is not competitive when compared with the optimal oracle scheduler:
	for any \(r > 0\), there exists an input sequence \(\sigma\) such that
	\begin{equation}
		\frac{\mathsf{Goodput}(\mathrm{OPT})}{\mathsf{Goodput}(\mathrm{EDF})}
		\;>\;
		r
	\end{equation}
\end{theorem}

\begin{proof}
	We construct an input sequence \(\sigma\) consisting of multiple requests. Let $T > 0$ be a fixed time, and let $N$ be a large positive integer.
	Define $\delta = \displaystyle \frac{T}{N + 1}$. The request sequence includes:
	\begin{itemize}[leftmargin=*]
		\item One request $A$ that arrives at time $0$, with computing time $t_{\mathrm{comp}}(A) = T$ and SLO time $t_{\mathrm{SLO}}(A) = T$, and goodput $R(A) = M$, where $M$ is a large positive number to be chosen later.
		\item $N$ requests $B_i$ for $i = 0, 1, \cdots, N - 1$ , each arriving at time $t_i = i \cdot \delta$, with computing time $t_{\mathrm{comp}}(B_i) = \delta$ and SLO time $t_{\mathrm{SLO}}(B_i) = T + \delta$, and goodput $R(B_i) = 1$.
	\end{itemize}
	\paragraph{EDF scheduling.}
	At time \(0\), request \(A\) arrives.
	Almost immediately, request \(B_0\) arrives at time \(0\) with SLO time \(\delta < T\), so EDF preempts \(A\) to serve \(B_0\), which completes at time \(\delta\).
	At time \(\delta\), request \(B_1\) arrives with SLO time \(2 \cdot \delta < T\), so EDF preempts \(A\) to serve \(B_1\), which completes at time \(2 \cdot \delta\).
	This process continues: at each time \(i \cdot \delta\), EDF schedules \(B_i\), which completes at time \((i + 1) \cdot \delta\).
	The last request \(B_{N - 1}\) is scheduled at time \((N - 1) \cdot \delta\) and completes at time \(N \cdot \delta\).
	At time \(N \cdot \delta\), no more \(B\) requests are available. EDF then schedules request \(A\). However, the current time is \(\displaystyle N \cdot \delta = \frac{N \cdot T}{N + 1} < T\), and \(A\) requires computing time \(T\).
	Thus, \(A\) completes at time \(N \cdot \delta + T > T\), missing its SLO.
	Therefore, \(A\) contributes \(0\) goodput.
	And the total goodput for EDF is the sum of goodputs from all \(B\) requests: \(\mathsf{Goodput}(\mathrm{EDF}) = N \times 1 = N\).

	\paragraph{Optimal scheduling.}
	It is clear to see that OPT will ignore all \(B_i\) requests and schedule request \(A\) starting at time \(0\).
	Since \(t_\mathrm{comp}(A) = T\) and its SLO is at time \(T\), \(A\) completes exactly at time \(T\), yielding \(\mathsf{Goodput}(\mathrm{OPT}) = M\).

	\paragraph{Competitive ratio.}
	Combining two cases above, we have the inverted competitive ratio
	\begin{equation}
		\frac{\mathsf{Goodput}(\mathrm{OPT})}{\mathsf{Goodput}(\mathrm{EDF})}
		\;=\;
		\frac{M}{N}
	\end{equation}
	For any \(r > 0\), choose \(\displaystyle M > \frac{N}{r}\). Then \(\displaystyle \frac{M}{N} > r\), proving the theorem.
\end{proof}

\paragraph{Remark.}
The classical Earliest Deadline First (EDF) policy is goodput-agnostic and therefore susceptible to adversarial workload constructions.
In particular, an attacker can inject a stream of low-goodput requests whose \emph{deadlines (SLOs)} are only marginally earlier than those of high-value jobs.
EDF will systematically favor these low-value, tight-SLO requests, repeatedly preempting or delaying lucrative work and thereby degrading system-level utility.

\subsubsection{Analysis of Shortest Job First Scheduling}
\label{sec:SJF}
\begin{theorem}[Non-competitiveness of SJF]
	The scheduling of Shortest Job First (SJF) is not competitive when compared with the optimal oracle scheduler:
	for any \(r > 0\), there exists an input sequence \(\sigma\) such that
	\begin{equation}
		\frac{\mathsf{Goodput}(\mathrm{OPT})}{\mathsf{Goodput}(\mathrm{SJF})}
		\;>\;
		r
	\end{equation}
\end{theorem}
\begin{proof}
	We consturct an input sequence \(\sigma\) consisting of multiple requests.
	Let $T > 0$ be a fixed time, and let $N$ be a large positive integer.
	Define $\delta = \frac{T}{N + 1}$.
	The sequence includes:
	\begin{itemize}[leftmargin=*]
		\item One request A that arrives at time \(0\), with computing time \(t_{\mathrm{comp}}(A) = T\) and SLO time \(t_{\mathrm{SLO}}(A) = T\), and goodput \(R(A) = M\), where \(M\) is a large positive number to be chosen later.
		\item N requests \(B_i\) for \(i = 0, 1, \cdots, N - 1\), each arriving at time \(t_i = i \cdot \delta\), with computing time \(t_{\mathrm{comp}}(B_i) = \delta\) and SLO time \(t_{\mathrm{SLO}}(B_i) = \delta\), and goodput \(R(B_i) = 1\).
	\end{itemize}

	\paragraph{SJF scheduling.}
	At time \(0\), request \(A\) arrives.
	Almost immediately, request \(B_0\) arrives at time \(0\) with computing time \(\delta < T\), so SJF preempts \(A\) to serve \(B_0\), which completes at time \(\delta\).
	At time \(\delta\), request \(B_1\) arrives with computing time \(\delta < T\), so SJF serves \(B_1\), which completes at time \(2 \cdot \delta\).
	This process continues: at each time \(i \cdot \delta\), SJF schedules \(B_i\), which completes at time \((i + 1) \cdot \delta\).
	At time \(N \cdot \delta\), no more \(B\) requests are available.
	SJF then schedules request \(A\).
	However, the current time is \(\displaystyle N \cdot \delta = \frac{N \cdot T}{N + 1} < T\), and \(A\) requires computing time \(T\).
	Thus, request \(A\) completes at time \(N \cdot \delta + T > T\), missing its SLO.
	Therefore, request \(A\) contributes \(0\) goodput.

	\paragraph{Optimal scheduling.}
	OPT will ignore all \(B_i\) requests and schedule request \(A\) starting at time \(0\).
	Since \(t_\mathrm{comp}(A) = T\) and its SLO is at time \(T\), request \(A\) completes exactly at time \(T\), yielding \(\mathsf{Goodput}(\mathrm{OPT}) = M\).

	\paragraph{Competitive ratio.}
	Combining two cases above, we have the inverted competitive ratio
	\begin{equation}
		\frac{\mathsf{Goodput}(\mathrm{OPT})}{\mathsf{Goodput}(\mathrm{SJF})}
		\;=\;
		\frac{M}{N}
	\end{equation}
	For any \(r > 0\), choose \(\displaystyle M > \frac{N}{r}\). Then \(\displaystyle \frac{M}{N} > r\), proving the theorem.
\end{proof}

\paragraph{Remark.}
Similarly, Shortest Job First (SJF) is indifferent to goodput and can be exploited by workloads populated with many low-goodput jobs of slightly shorter \emph{computing times}.
SJF will preferentially execute these short, low-value tasks, crowding out requests that are only marginally longer yet yield much higher goodputs, which leads to poor aggregate performance.

\subsection{Analysis of \name Scheduling}
\label{app:JITServe}
We set the following condition as our additional preemption threshold:
a newly considered request \(A\) may preempt the currently running request \(B\) if
\begin{equation}
	\frac{R(A)}{R(B)}
	\;>\;
	1 + \delta
	\qquad \text{for} \; \delta > 0
\end{equation}
We set this preemption threshold to avoid possible preemption overhead, and only a request with higher goodput may interrupt the current request.

\begin{lemma}[Constant competitiveness of \name without GMAX]
\label{lemma:no-gmax}
	The scheduling of \name without GMAX is constant competitive when compared with the optimal oracle scheduler:
	there exists \(r > 0\) such that
	\begin{equation}
		\frac{\mathsf{Goodput}(\name)}{\mathsf{Goodput}(\mathrm{OPT})}
		\;\ge\;
		r
	\end{equation}
\end{lemma}

\begin{proof}
	We use standard competitive analysis to evaluate our scheduling algorithm.
	And we will proceed our proof by using the \textit{credit charging} technique from the amortized analysis.
	We first map the goodput of OPT to the requests served by \name via a carefully designed fractional credit charging mapping that respects their relative time overlap.
	Then we use the preemption-chain amortization to bound the total goodput of \name by the goodput of completed requests.
	And finally we combine the two pieces together to get the final competitive ratio.

	\paragraph{Charging credits.}
	We charge credits from each request $U \in \sigma_S$ to its corresponding request $V \in \sigma_*$, and we denote this credit charging rule as $f(U, V)$.
	In our setting, the system’s service capacity $\mathfrak{C}$ is modeled as $\mathfrak{C}$ parallel service slots.
	For expositional clarity, we first analyze the single slot case, i.e., schedules with $|R| =1$.
	The extension to multiple slots is immediate, since the analysis applies independently to each slot.
	To better differentiate symbols and notations, we use $U$ to denote the request in \name's schedule $\sigma_S$, and $V$ to denote the request in the optimal schedule $\sigma_*$.
	Without loss of generality, we define the properties of a \emph{chargable credit mapping} as follows:
	\noindent
	\begin{itemize}[leftmargin=*]
		\item \textbf{Property 1:} For any request $U \in \sigma_S$, its aggregated charged credits to all requests in $\sigma_*$ should be equal to the goodput of $U$, i.e., we have:
		      \begin{equation}
			      \sum_{V \in \sigma_*} f(U, V) = R(U), \quad \forall U \in \sigma_S
		      \end{equation}
		\item \textbf{Property 2:} The aggregated charged credit from all $U \in \sigma^c_S$ to all $V \in \sigma_*$ must be greater than or equal to a constant portion of the aggregated goodput of $V \in \sigma_*$:
		      \begin{equation}
			      \sum_{V \in \sigma_*} \sum_{U \in \sigma_S^c} f(U, V) \ge r\sum_{V \in \sigma_*} R(V)
		      \end{equation}
		      where $r \in [0, 1]$ is a constant.
	\end{itemize}

	\noindent
	Note that for any chargable credit mapping with the above two properties, we have the following lemma:
	\begin{lemma}
		\label{lemma:chargable_mapping}
		For any chargable credit mapping \(f: (U, V) \rightarrow \mathbb{R}\) with a constant portion factor \(r\) in \textbf{Property 2}, we have:
		\begin{equation}
			\frac{\mathsf{Goodput}(\sigma_S)}{\mathsf{Goodput}(\sigma_*)} \ge r
		\end{equation}
	\end{lemma}
	\begin{proof}[Proof of lemma]
		\begin{align*}
			\mathsf{Goodput}(\sigma_S) & = \sum_{U \in \sigma_S^c} R(U)                          \\
			                          & = \sum_{U \in \sigma_S^c} \sum_{V \in \sigma_*} f(U, V) \\
			                          & = \sum_{V \in \sigma_*} \sum_{U \in \sigma_S^c} f(U, V) \\
			                          & \ge \sum_{V \in \sigma_*} r R(V)                        \\
			                          & = r \sum_{V \in \sigma_*} R(V)                          \\
			                          & \ge r \cdot \mathsf{Goodput}(\sigma_*)
		\end{align*}
	\end{proof}

	The key challenge here lies in constructing such a chargable credit mapping and quantifying such a good constant \(r\).

	With lemma \ref{lemma:chargable_mapping}, we are now ready define such credit charging mapping, which assigns to each ordered pair \((U,V)\) a nonnegative charged credit \(f(U,V)\) according to the following rules:
	\begin{itemize}[leftmargin=*]
		\item \textbf{Rule 1:} If \(\displaystyle s_S(V) \le s_*(U)\), and \(\displaystyle t^\mathrm{r}_\mathrm{comp}(V) > t^\mathrm{r}_\mathrm{comp}(U)\), we set $\displaystyle f(U,V) \coloneqq \alpha \cdot R(U)$.
		\item \textbf{Rule 2:} If \(\displaystyle s_S(V) \le s_*(U)\), and \(\displaystyle t^\mathrm{r}_\mathrm{comp}(V) \le t^\mathrm{r}_\mathrm{comp}(U)\), we set $\displaystyle f(U,V) \coloneqq \beta \cdot \frac{t^\mathrm{r}_\mathrm{comp}(V) + \epsilon}{t^\mathrm{r}_\mathrm{comp}(U) + \epsilon} \cdot R(U)$.
		\item \textbf{Rule 3:} If \(V\) and \(U\) share the same request, we set $\displaystyle f(U,V) \coloneqq \gamma \cdot (1 + \delta)^3 \cdot R(U)$.
		\item \textbf{Rule 4:} Finally, if the aggregated credit mapped to $V$ is less than $R(U)$, we assign the residual $R(U)$ to any arbitrary $V \in \sigma_*$.
	\end{itemize}
	It's clear to see that if we want this credit charging mapping $f$ to be a \emph{chargable} credit mapping, we will need to fix these three constants \(\alpha \ge 0, \beta \ge 0, \gamma \ge 0\) with an additional condition that \(\alpha + \beta + \gamma \le 1\).

	\begin{lemma}
		The credit charging mapping $f$ satisties the above \textbf{Property 1} of a chargable credit mapping.
	\end{lemma}
	\begin{proof}[Proof of lemma]
		For now, we assume that our credit charging mapping is confined to the above Rule 1, 2, and 3, and we have:
		\begin{align*}
			    & \;\; \sum_{V \in \sigma_*} \, g(U, V)                         \\
			=   & \quad \sum_{\substack{V: s_*(V) \ge s_S(U)                    \\ t_\mathrm{SLO}(V) > t_\mathrm{SLO}(U)}} \alpha \cdot R(U) \\
			    & + \sum_{\substack{V: s_*(V) \ge s_S(U)                        \\ t_\mathrm{SLO}(V) \le t_\mathrm{SLO}(U)}} \beta \cdot \frac{t^\mathrm{r}_\mathrm{comp}(V) + \epsilon}{t^\mathrm{r}_\mathrm{comp}(U) + \epsilon} \cdot R(U) \\
			    & + \quad \;\;\, \sum_{V: V = U} \quad \;\;\; \gamma \cdot R(U) \\
			\le & \quad \sum_{\substack{V: s_*(V) \ge s_S(U)                    \\ t_\mathrm{SLO}(V) > t_\mathrm{SLO}(U)}} \alpha \cdot R(U) \;\; + \sum_{\substack{V: s_*(V) \ge s_S(U)                          \\ t_\mathrm{SLO}(V) \le t_\mathrm{SLO}(U)}} \beta \cdot R(U) \\
			    & + \quad \;\;\, \sum_{V: V = U} \quad \;\;\; \gamma \cdot R(U) \\
			=   & \;\; (\alpha + \beta + \gamma) \cdot R(U)                     \\
			\le & \;\;\, R(U)
		\end{align*}
		To ensure \textbf{Property 1}, we can simply follow the Rule 4 to assign the residual credit to any arbitrary \(V\).
		Therefore, we have:
		\begin{align*}
			\sum_{V \in \sigma_*} f(U, V) & = \sum_{V \in \sigma_*} \Bigl( g(U, V) + h(U, V) \Bigr)                           \\
			                              & = \sum_{V \in \sigma_*} g(U, V) +  \sum_{V \in \sigma_*} h(U, V)                  \\
			                              & = (\alpha + \beta + \gamma) \cdot R(U) + (1 - \alpha - \beta - \gamma) \cdot R(U) \\
			                              & = R(U)
		\end{align*}
		where \(h(U, V) \ge 0\) denotes the residual credit assigned by Rule 4 in the credit mapping $f$.
	\end{proof}

	\paragraph{Per-rule lower bounds.}
	The remaining tasks now are all about how to bound a good constant \(r\) in \textbf{Property 2}.
	Fortunately, with the \emph{credit charging} mapping technique, we can now focus on bounding the credit contribution to \(V\) from each \(U \in \sigma_S\).
	Note that there's a brand-new challenge to address: there are two possible reasons why \(V\) may not be running at time \(s_*(V)\).
	As a concrete example, consider the case where \(U\) starts at time \(t_U\) and \(V\) starts at time \(t_V = t_U + \epsilon > t_U\).
	On the one hand, \(V\) may be unable to preempt the request because \(U\) that is currently running.
	On the other hand, request \(V\) may have already ended in \(\sigma_S\) so there is no way for \(V\) to be served again in \(\sigma_S\).
	We denote the unfinished request \(V\) as \(V^U\), and the completed request \(V\) as \(V^C\).

	\paragraph{\(V^U\) Analysis}
	In \(V^U\) analysis, we only consider the credit charging Rule 1 and 2.
	If \(V^U\) starts later than some request \(U\), then \(V^U\) must be blocked by request \(U\), hence we have the following two cases.

	\paragraph{Case 1:} We have \(\displaystyle s_S(V) \le s_*(U)\) and \(\displaystyle t^\mathrm{r}_\mathrm{comp}(V) > t^\mathrm{r}_\mathrm{comp}(U)\).
	And therefore Rule 1 can be applied.
	According to the preemption threshold, we have two possible cases:
	\begin{itemize}[leftmargin=*]
		\item \textbf{Case 1a:} \(V\) is not served because
		      \begin{equation}
			      \label{eq:case1a}
			      \frac{R(V)}{t^\mathrm{r}_\mathrm{comp}(V) + \epsilon}
			      \; \le \;
			      \frac{R(U)}{t^\mathrm{r}_\mathrm{comp}(U) + \epsilon}
		      \end{equation}
		      Conbined with \textbf{Rule 1} and Eq. \ref{eq:case1a}, we have:
		      \begin{align}
			      \sum_{U \in \sigma_S} f(U, V) & \ge \alpha \cdot R(U)                                                                                                                                         \\
			                                    & \ge \alpha \cdot \frac{t^\mathrm{r}_\mathrm{comp}(U) + \epsilon}{t^\mathrm{r}_\mathrm{comp}(V) + \epsilon} \cdot R(V) \\
			                                    & \ge \alpha \cdot R(V) \label{eq:case1a_bound}
		      \end{align}
		\item \textbf{Case 1b:} \(V\) is not served because
		      \begin{equation}
			      \label{eq:case1b}
			      \frac{R(V)}{R(U)} \le 1 + \delta
		      \end{equation}
		      Combined with \textbf{Rule 1} and Eq. \ref{eq:case1b}, we have:
		      \begin{align}
			      \sum_{U \in \sigma_S} f(U, V) & \ge \alpha \cdot R(U)                                            \\
			                                    & \ge \alpha \cdot \frac{R(V)}{1 + \delta} \label{eq:case1b_bound}
		      \end{align}
	\end{itemize}

	\paragraph{Case 2:} We have \(\displaystyle s_S(V) \le s_*(U)\), and \(\displaystyle t_\mathrm{SLO}(V) \le t_\mathrm{SLO}(U)\).
	And therefore Rule 2 can be applied.
	According to the preemption threshold, we have two possible cases:
	\begin{itemize}[leftmargin=*]
		\item \textbf{Case 2a:} \(V\) is not served because
		      \begin{equation}
			      \label{eq:case2a}
			      \frac{R(V)}{t^\mathrm{r}_\mathrm{comp}(V) + \epsilon}
			      \; \le \;
			      \frac{R(U)}{t^\mathrm{r}_\mathrm{comp}(U) + \epsilon}
		      \end{equation}
		      Conbined with \textbf{Rule 2} and Eq. \ref{eq:case2a}, we have:
		      \begin{align}
			      \sum_{U \in \sigma_S} f(U, V) & \ge \beta \cdot \frac{t^\mathrm{r}_\mathrm{comp}(V) + \epsilon}{t^\mathrm{r}_\mathrm{comp}(U) + \epsilon} \cdot R(U)            \\
			                                    & \ge \beta \cdot \left( \frac{t^\mathrm{r}_\mathrm{comp}(V) + \epsilon}{t^\mathrm{r}_\mathrm{comp}(U) + \epsilon} \right)        \\
			                                    & \quad \cdot \left( \frac{t^\mathrm{r}_\mathrm{comp}(U) + \epsilon}{t^\mathrm{r}_\mathrm{comp}(V) + \epsilon} \right) \cdot R(V) \\
			                                    & \ge \beta \cdot R(V) \label{eq:case2a_bound}
		      \end{align}

		\item \textbf{Case 2b:} \(V\) is not served because
		      \begin{equation}
			      \label{eq:case2b}
			      \displaystyle \frac{R(V)}{R(U)} \le 1 + \delta
		      \end{equation}
		      Combined with \textbf{Rule 2} and Eq. \ref{eq:case2b}, we have:
		      \begin{align}
			      \sum_{U \in \sigma_S} f(U, V) & \ge \beta \cdot \frac{t^\mathrm{r}_\mathrm{comp}(V) + \epsilon}{t^\mathrm{r}_\mathrm{comp}(U) + \epsilon} \cdot R(U) \\
			                                    & \ge \beta \cdot R(U)                                                                                                                                         \\
			                                    & \ge \beta \cdot \frac{R(V)}{1 + \delta} \label{eq:case2b_bound}
		      \end{align}
	\end{itemize}

	\paragraph{\(V^C\) Analysis}
	Note that \(V^C\) must have been completed in \(\sigma_S\). So based on the charging Rule 3, any request \(V\) must have been charged with a value of $\gamma \cdot R(U)$.
	Therefore, we have:
	\begin{align}
		\sum_{U \in \sigma_S} f(U, V) & \ge \gamma \cdot (1 + \delta)^3 \cdot R(U) \\
		                              & = \gamma \cdot (1 + \delta)^3 \cdot R(V)
	\end{align}

	\noindent
	Combining all the above cases, i.e., Eqs. \ref{eq:case1a_bound}, \ref{eq:case1b_bound}, \ref{eq:case2a_bound}, and \ref{eq:case2b_bound}, we have the following inequality:
	\begin{equation}
		\label{eq:per_request_bound}
		\sum_{U \in \sigma_S} f(U, V)
		\ge
		\min
		\Bigl(
		\alpha, \,
		\frac{\alpha}{1 + \delta}, \,
		\beta, \,
		\frac{\beta}{1 + \delta}, \,
		\gamma \cdot (1 + \delta)^3
		\Bigr)
		\cdot R(V)
	\end{equation}

	\paragraph{Preemption chains.}
	Based on the previous preemption threshold, we can now bound the total goodput of requests served by \name by the goodput of completed requests.
	Note that we can now partition the requests served by \name into chains:
	\[
		U_1 \prec U_2 \prec \cdots \prec U_m,
	\]
	where \(U_{k+1}\) directly preempts \(U_k\).
	By the preemption threshold, we have
	\begin{equation}
		R(U_{k+1}) \; > \; (1 + \delta) \cdot R(U_k)
	\end{equation}
	hence by the property of geometric series,
	\begin{equation}
		\sum_{k=1}^{n} R(U_k) \le \frac{1 + \delta}{\delta} \cdot R(U_n)
	\end{equation}
	Moreover, every chain terminates at a request \(U_n \in \sigma_S^c\) (the last request in the chain completes on time).
	Summing over chains yields:
	\begin{align}
		\label{eq:preemption_chain}
		\sum_{U\in\sigma_S} R(U) & \le \frac{1 + \delta}{\delta} \sum_{I\in\sigma_S^c} R(I)    \\
		                         & = \frac{1 + \delta}{\delta} \cdot \mathsf{Goodput}(\sigma_S)
	\end{align}

	\paragraph{Amortized charging bound.}
	With Eq. \ref{eq:preemption_chain}, we have:
	\begin{align}
		\mathsf{Goodput}(\sigma_S) & \ge \frac{\delta}{1 + \delta} \cdot \sum_{U \in \sigma_S} R(U)                        \\
		                          & = \frac{\delta}{1 + \delta} \cdot \sum_{U \in \sigma_S} \sum_{V \in \sigma_*} f(U, V) \\
		                          & = \frac{\delta}{1 + \delta} \cdot \sum_{V \in \sigma_*} \sum_{U \in \sigma_S} f(U, V)
	\end{align}
	Combining with the per-request bound Eq. \ref{eq:per_request_bound}, we have:
	\begin{align}
		\mathsf{Goodput}(\sigma_S) & \ge \frac{\delta}{1 + \delta} \cdot \mathfrak{A}(\delta, \alpha ,\beta, \gamma) \cdot \sum_{V \in \sigma_*} R(V) \\
		                          & = \frac{\delta}{1 + \delta} \cdot \mathfrak{A}(\delta, \alpha ,\beta, \gamma) \cdot \mathsf{Goodput}(\sigma_*)
	\end{align}
	where
	\begin{equation}
		\mathfrak{A}(\delta, \alpha ,\beta, \gamma)
		\;=\;
		\min
		\Bigl(
		\alpha, \,
		\frac{\alpha}{1 + \delta}, \,
		\beta, \,
		\frac{\beta}{1 + \delta}, \,
		\gamma \cdot (1 + \delta)^3
		\Bigr)
	\end{equation}
    
	Therefore, we can conclude that there exists a constant competitive ratio for \name:
	\begin{equation}
    \label{eq:bound-lemma}
		\mathfrak{B}(\delta, \alpha ,\beta, \gamma)
		\;=\;
		\frac{\delta}{1 + \delta} \cdot
		\min
		\Bigl(
		\frac{\alpha}{1 + \delta},
		\frac{\beta}{1 + \delta}, \,
		\gamma \cdot (1 + \delta)^3
		\Bigr)
	\end{equation}

	\paragraph{Optimize objective function.}
	It is clear that the maximum lower bound can be achieved by optimizing over \(\delta, \alpha, \beta, \gamma\)
	under the constraints \(\delta>0\), \(\alpha \ge 0,\beta \ge 0, \gamma \ge 0\), and \(\alpha + \beta + \gamma \le1\), i.e.,
	formally, we need to solve the optimization problem to attain the maximum competitive ratio:
	\begin{align}
		\max_{\delta,\, \alpha,\, \beta,\, \gamma} \quad & \mathfrak{B}(\delta, \alpha, \beta, \gamma) \\
		\mathsf{s.t.}  \quad                             & \alpha + \beta + \gamma \le 1               \\
		                                                 & \alpha \ge 0,\, \beta \ge 0, \gamma \ge 0   \\
		                                                 & \delta > 0
	\end{align}

    \begin{figure}[H]
        \centering
        \includegraphics[width=0.8\linewidth]{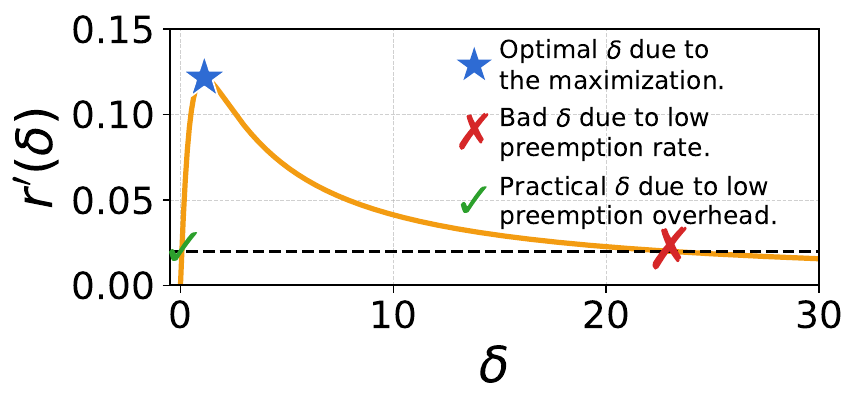}
        \caption{Competitive ratio \(r'(\delta)\) versus preemption threshold \(\delta\)}
        \label{fig:pmtn-overhead}
    \end{figure}
    
Solving these above optimization problem using \emph{numerical analysis}~\cite{numerical-analysis} yields an optimal performance guarantee of \name $r'(\delta) \approx \frac{1}{8.13}$. Note that as we discussed in Section~\ref{sec:design}, preemption can improve the performance of our online algorithm \name, but it also introduces practical overhead. Figure~\ref{fig:pmtn-overhead} illustrates the trade-off between preemption overhead and the competitive ratio: the $x$-axis is the preemption threshold $\delta$, and the $y$-axis is the performance guarantee of \name. To balance objective maximization and implementation overhead, we choose a moderate threshold of $\delta=10\%$, which slightly relaxes the bound yet yields a performance guarantee that remains acceptable in practice.

\end{proof}

\begin{theorem}[Constant competitiveness of \name with GMAX]
	The scheduling of \name with GMAX is constant competitive when compared with the optimal oracle scheduler:
	there exists \(r > 0\) such that
	\begin{equation}
		\frac{\mathsf{Goodput}(\name)}{\mathsf{Goodput}(\mathrm{OPT})}
		\;\ge\;
		r
	\end{equation}
\end{theorem}

\begin{proof}
We now continue from the above lemma to complete the proof. Note that we now have already had a performance guarantee for our \name scheduling algorighm under no-GMAX scenario. We are now ready to prove that our \name scheduling together with GMAX also has a performance guarantee.

\paragraph{Top-$p$ Filtering}
We are now ready to prove that GMAX will only introduce a uniform \((1-\varepsilon)\)-loss surrogate for each served request.
Fix any batching decision time.
Let \(R(r(b))\) denote the \(b\)-th largest goodput value among all currently available requests.
By the GMAX rule, the candidate set is
\begin{equation}
\mathcal{T}:=\{W:\ R(W)\ge p \cdot R(r(b))\}.
\end{equation}
Consider any request \(U\) that \(\sigma_S\) would serve (or continue to serve) at this time. Because \(\sigma_S\) maintains batch size \(b\), necessarily \(R(U)\ge R(r(b))\).
Hence there exists some \(U'\in \mathcal{T}\) with
\begin{equation}
\label{eq:surrogate-loss}
R(U')\ \ge\ p \,R(r(b))\ \ge\ p\,R(U).
\end{equation}
Intuitively, \(U'\) is a \emph{surrogate} for \(U\) that is always almost as valuable in terms of its goodput value; the length adjacency constraint in \textsf{GMAX} only determines \emph{which} \(b\) requests inside \(\mathcal{T}\) are taken, but it never drives any selected \(R(\cdot)\) below the threshold \( p \cdot R(r(b))\).
Therefore, replacing \(U\) by \(U'\) in any lower-bound argument causes at most a multiplicative $p$ degradation.

\paragraph{Putting together.}
Combining Eq.~\ref{eq:bound-lemma} from Lemma~\ref{lemma:no-gmax} and Eq.~\ref{eq:surrogate-loss}, we now have
\begin{equation}
 \mathsf{Goodput}(\sigma_S) 
\ge \frac{p \cdot \delta}{1+\delta} \cdot 
\mathfrak{B}(\delta, \alpha, \beta, \gamma) \cdot \mathsf{Goodput}(\sigma_*)
\end{equation}
Solving these above optimization problem using \emph{numerical analysis}~\cite{numerical-analysis} under our experimental setting yields an optimal performance guarantee of \name $r(\delta) \approx \frac{1}{8.557}$. 

\end{proof}

\end{document}